\documentclass[graybox]{svmult}

\def\uu {4U\,0142$+$614}
\def\ee {1E\,1048.1$-$5937}
\def\kes {1E\,1841$-$045}
\def\aa {1E\,1547$-$5408}
\def\axj {AX\,J1844$-$0258}
\def\rxs {1RXS\,J1708$-$4009}
\def\xte{XTE\,J1810$-$197\,}
\def\cxo{CXOU\,J0100$-$7211\,}
\def\wes{CXOU\,J1647$-$4552}
\def\ea {1E\,2259$+$586}
\def\hess{CXOU\,J1714$-$3810}
\def\sgra{SGR\,1806$-$20}
\def\sgrb{SGR\,1900$+$14}
\def\sgrc{SGR\,0526$-$66\,}
\def\sgrd{SGR\,1627$-$41\,}
\def\sgre{SGR\,0501$+$4516}
\def\sgrf{SGR\,0418$+$5729}
\def\sgrg{SGR\,1833$-$0832}

\def\psr{PSR\,1622$-$4950}
\def\hbpsr{PSR\,J1846$-$0258}
\newcommand{\CXO}{Chandra\,}

\newcommand{\xmm}{XMM-Newton}

\newcommand{\su}{Suzaku}

\newcommand{\swift}{Swift}

\newcommand{\INT}{INTEGRAL}
\newcommand{\bc}{\begin{center}}
\newcommand{\ec}{\end{center}}
\def\ltsima{$\; \buildrel < \over \sim \;$}
\def\lsim{\lower.5ex\hbox{\ltsima}}
\def\loe{\lower.5ex\hbox{\ltsima}}
\def\gtsima{$\; \buildrel > \over \sim \;$}
\def\gsim{\lower.5ex\hbox{\gtsima}}
\def\goe{\lower.5ex\hbox{\gtsima}}

\def\ltsima{$\; \buildrel < \over \sim \;$}
\def\lsim{\lower.5ex\hbox{\ltsima}}
\def\loe{\lower.5ex\hbox{\ltsima}}
\def\gtsima{$\; \buildrel > \over \sim \;$}
\def\gsim{\lower.5ex\hbox{\gtsima}}
\def\goe{\lower.5ex\hbox{\gtsima}}

\def\ergs {erg\,s$^{-1}$}
\def\ergscm2 {erg\,s$^{-1}$cm$^{-2}$}
\def\ss {s\,s$^{-1}$}
\def\cm2 {cm$^{-2}$}

\usepackage[numbers]{natbib}

\usepackage{aas_macros}
\usepackage{mathptmx}       
\usepackage{helvet}         
\usepackage{courier}        
\usepackage{type1cm}        
\usepackage{makeidx}         
\usepackage{graphicx}        
\usepackage{multicol}        
\usepackage[bottom]{footmisc}

\makeindex             

\begin{document}

\title*{Magnetar outbursts: an observational review}
\author{Nanda Rea \& Paolo Esposito}
\institute{Nanda Rea \at Institut de Ciences de l'Espai (CSIC-IEEC), Campus UAB, Facultat de Ciencies, Torre C5-parell 2a planta, Bellaterra (Barcelona), Spain; \email{rea@ice.csic.es} \and Paolo Esposito \at Osservatorio Astronomico di Cagliari, localit\`a Poggio dei Pini, strada 54, 09012, Capoterra, Italy; \email{paoloesp@oa-cagliari.inaf.it}}
\authorrunning{Nanda Rea \& Paolo Esposito}
\maketitle

\abstract{Transient outbursts from magnetars have shown to be a key property of their emission, and one of the main way to discover new sources of this class. From the discovery of the first transient event around 2003, we now count about a dozen of outbursts, which increased the number of these strongly magnetic neutron stars by a third in six years.  Magnetars' outbursts might involve their multi-band emission resulting in an increased activity from radio to hard X-ray, usually with a soft X-ray flux increasing by a factor of 10--1000 with respect to the quiescent level. A connected X-ray spectral evolution is also often observed, with a spectral softening during the outburst decay. The flux decay times vary a lot from source to source, ranging from a few weeks to several years, as also the decay law which can be exponential-like, a power-law or even multiple power-laws can be required to model the flux decrease. We review here on the latest observational results on the multi-band emission of magnetars, and summarize one by one all the transient events which could be studied to date from these sources.}

\section{Authors' preface}
\label{preface}

The magnetar field have been recently boosted by the discovery of transients magnetars, and more in general by their possible role in gamma-ray bursts and gravitational wave researches. However, probably ``because" of the rapid development of the field, there is still a large confusion in the literature on when a source can be labelled as a magnetar candidate, and what exactly this word means: is the super-critical dipolar magnetic field which defines a magnetar? Is the bursting behavior? Is the low rotational power with respect to their X-ray luminosity? Is the blackbody plus power-law X-ray spectrum? Is the erratic radio pulsed behavior? When do we define a source an Anomalous X-ray Pulsar (AXP) or a Soft Gamma Repeater (SGR)? Apparently in the recent literature, even groups working in the magnetar field since decades are not fully in agreement with their exact definition: see the archetypical example of the i) AXP \aa\ = SGR 1550--5418 = PSR J1550--5418 which has been discovered only 3 years ago and it already has three names, ii) \hbpsr\, which showed all the typical magnetar-like activity but it is still labelled as a fully rotational powered pulsar since as such it was first discovered, or iii) \sgrf\, which shows all the typical magnetar emission properties but it has a magnetic field in line with normal pulsars. Most of the questions above can be answered with a counter-example, it is then indeed becoming very difficult to give a definitive and unique answer on when we can call a source an AXP, an SGR, or a magnetar in general.

With this preface we aim at warning non-expert readers on the assumption and conceptual choices we will make in this review. In particular, this is a pure observational review, mainly focussed on transients. We will then enter very little in the theoretical interpretations. We will not discuss about X-ray bursts or flares, but only about the outbursts of the persistent emission of these sources. Furthermore, a part from the historical section, we will consider hereafter AXPs and SGRs as the same class of sources, calling them ``magnetar candidates". We apologize in advance if from time to time we drop the ``candidate" label in the text, but this is only due an easier writing. We will include \psr\, and \hbpsr\, in the magnetar list, given the discovery of their magnetars-like behavior, which clearly shows that (at least occasionally) they cannot be only rotational powered.

This review is structured as follow: a brief historical overview followed by the description of the multi-band emission of magnetars (with detailed numbers reported in the two tables rather than in the text). Then, we report one by one on all magnetars' outburst observed to date.

\section{A bit of history}

Neutron stars are the debris of the supernova explosion of massive stars, the existence of which was first theoretically
predicted around 1930 \citep{chandrasekhar31,bz34} and
then observed for the first time more than 30 years later \citep{hewish68}. We now know many different flavors of these compact
objects, and many open questions are still waiting for an answer after
decades of studies. The neutron star world is mainly populated by the radio
pulsars and the binary pulsars (thousands of objects), however in the
last decades also extreme and puzzling small sub-classes of neutron stars were
discovered: Anomalous X-ray Pulsars (AXPs), Soft Gamma Repeaters
(SGRs), Rotating Radio Transients (RRATs), X-ray Dim Isolated Neutron
Stars (XDINs), and Central Compact Objects (CCOs). The large amount of
different acronyms might already show how diverse is the neutron star
class, and on the other hand, how far we are from a unified
scenario. In particular, despite being presumably governed by a single
equation of state, the neutron star zoo manifests itself as a puzzling
multicolored class, whose bewildering variety of observational
properties is still largely unexplained. These objects are amongst the
most intriguing populations in modern high-energy astrophysics and in
physics in general. In fact, besides being interesting themselves in
terms of studying the neutron star equation of state and the physical processes
and mechanisms involved in their emission, they are precious places to
test gravitational and particle physics, relativistic plasma theories,
as well as strange quark states of matter and physics of atoms and
molecules embedded in extremely high magnetic fields (impossible to be
reproduced on Earth). The focus of this review is the outburst emission of strongly magnetized neutron stars, having magnetic fields close or stronger than the electron critical magnetic field of $B_{\rm
  crit} = m_{e}^2
c^3/e \hbar\sim4.4\times10^{13}$\,Gauss, at which the cyclotron energy of an
electron reaches the electron rest mass energy. 



\begin{figure}[t]
\includegraphics[width=12cm]{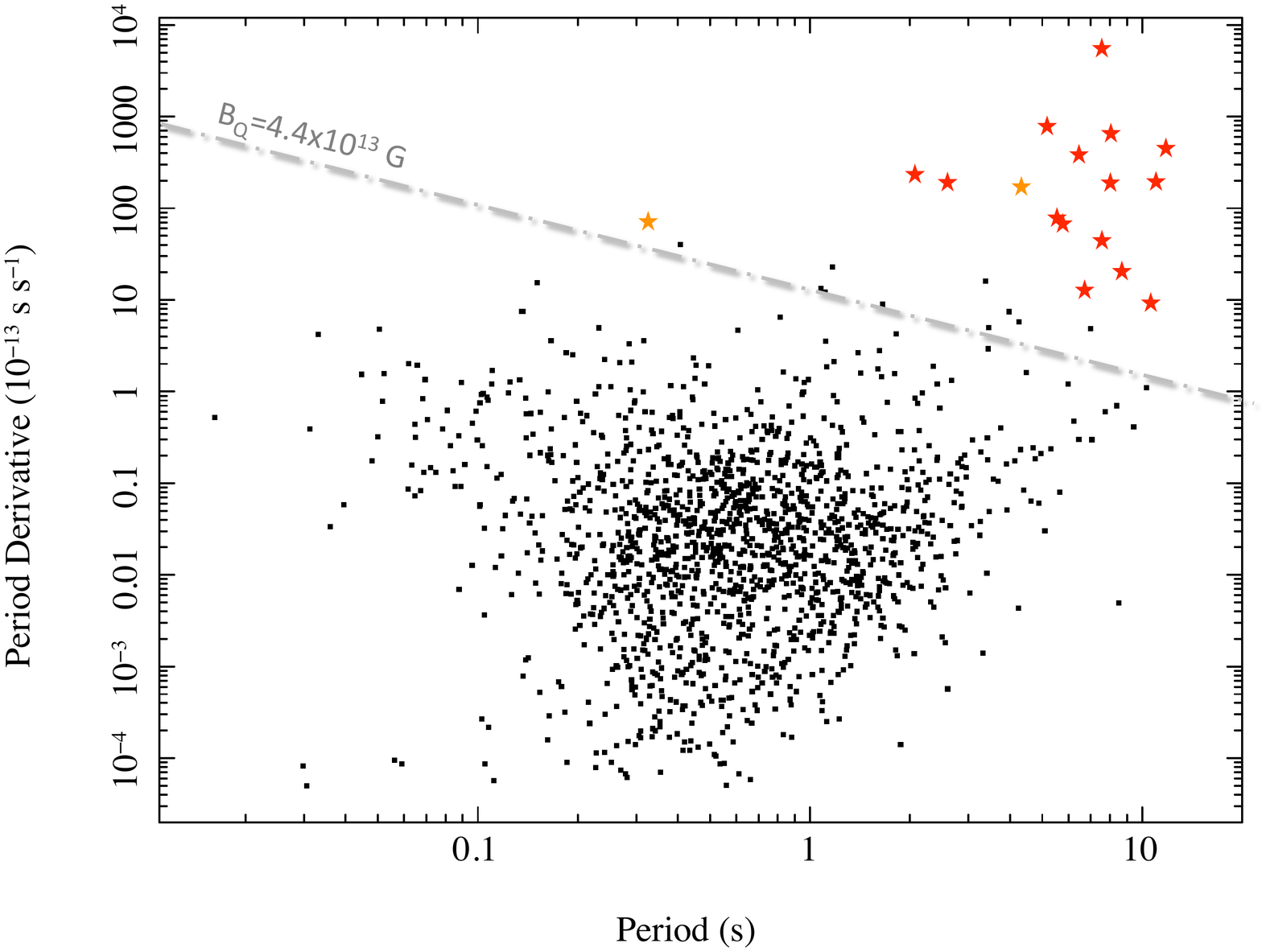}
\caption{Diagram of the $P$--$\dot{P}$ of all isolated pulsars known to date. Stars represents all sources which showed magnetar activity: in red all SGRs and AXPs having a measurement of the spin period and its derivative, and in orange \hbpsr\, and \psr .}
\label{ppdot}     
\end{figure}


\section{General observational characteristics}

Before presenting these ultra-magnetic objects, it is instructive to
indicate how the magnetic field of isolated pulsars is
estimated. Assuming that the spin-down torque is due to magnetic
dipole radiation, the surface magnetic field can be estimated from the
measured pulsar spin period $P$ and its derivative $\dot{P}$, for each
pulsar: $$B_{\rm surface} = (3~I~c^3 \dot{P}~P/8\pi^2 R^6)^{1/2} \sim
3.2\times10^{19}(P \dot{P})^{1/2}~{\rm Gauss}$$ (where $P$ is in units
of seconds, and $I\sim10^{45}$~g~cm$^2$ and $R\sim10^6$~cm are the
assumed neutron star moment of inertia and radius). Presently there
are almost 1800 spin-down powered radio pulsars known, with periods
from about 1.5~ms to 8~s \citep{manchester04}, and on average they have
magnetic fields of $\sim10^{12}$~Gauss. However, beside the magnetar
candidates which are the topic of this review, there are also a
handful of radio pulsars and other newly discovered type of pulsars
having super critical magnetic fields ( $> B_{\rm crit}$): namely the
high--B pulsars, a few XDINSs and RRATs. We will not report on those
in this review, however we want to point out that they might represent
somehow quiescent or evolved magnetars. The ``magnetars'' are a small group of X-ray pulsars
(about twenty objects with spin periods between 2--12 s) the emission of which
is very hardly explained by any of the common scenario for the radio
pulsar or the X-ray binary pulsar populations \citep{mereghetti08}. In fact, the very
strong X-ray emission of these objects is too high and variable to be
fed by the rotational energy alone (as in the radio pulsars), and no
evidence for a companion star has been found so far in favor of any
accretion process (as in the X-ray binary systems). Moreover, roughly
assuming them being magnetic dipole radiator, their inferred magnetic
fields appear to be as high as $B\sim10^{14}-10^{15}$~Gauss,
definitely larger than the quantum electron critical magnetic field
limit $B_{\rm crit}$ .  Because of these high B fields, the emission of magnetars is thought to be powered by the decay and the instability of
these strong fields \citep{duncan92,thompson93}, but despite valuable recent theories \citep{thompson02,beloborodov07}  a complete physical interpretation of all different aspects of their emission is still
missing.  Very interestingly, the magnetars are also characterized by
catastrophic and peculiar X-ray bursting, flaring and outbursting
events where luminosities of the order of $10^{46}$ \ergs\ are
reached. In Tables 1 and 2 we list the main characteristics of magnetars, and
in the rest of this review we discuss one by one the multi-band
properties of these objects, finishing with reviewing their transient
activity as known to date.

%
\begin{table}[t]
\caption{Accurate positions and timing characteristics of magnetars. }
\label{tab:1}       
%
%
\begin{tabular}{p{2.8cm}p{2.0cm}p{2.0cm}p{1.0cm}p{1.0cm}p{1.0cm}p{1.0cm}}
\hline\noalign{\smallskip}
Magnetars & RA  & Dec  & $P$ & $\dot{P}_{-12}$ & B$_{14}$ & d$^b$\\
 &  (J2000) &  (J2000) &  (s) & $ (s/s)$  & (G) & (kpc)\\
\noalign{\smallskip}\svhline\noalign{\smallskip}
\ea $^o$ &  23 01 08.29	& +58 52 44.45	  & 6.98  & 0.5 & 0.6 & 3.0\\
 \uu $^o$ & 01 46 22.44	 & +61 45 03.3   & 8.69 & 2.0 & 1.3 & 3.0 \\
 \rxs & 17 08 46.87	&  -40 08 52.44	  & 10.99 & 24* & 4.7 & 3.8 \\
 \ee $^o$ & 10 50 07.14	&  -59 53 21.4	&  6.45 &  50* & 4.4 & 2.7 \\
 \kes & 18 41 19.34	 & -04 56 11.16	 & 11.77 & 41 & 7.1 & 7.0 \\
 \cxo & 01 00 43.14	 & -72 11 33.8	&  8.02 & 19 & 3.9 & 60\\
 \wes $^o$ & 16 47 10.2	&  -45 52 16.9 & 10.61 & 0.9* & 1.3 & 5.0 \\ 
 \xte $^o$ & 18 09 51.08	&  -19 43 51.74	 & 5.54 & 10* & 1.6 & 2.5 \\
 \aa $^o$ & 15 50 54.11	&  -54 18 23.7	 & 2.07 & 23 * & 2.2 & 4.0 \\
 \hess & 17 14 05.74 & -38 10 30.9 & 3.82 & 59 & 4.8 & 8.0 \\
 \sgra $^o$ & 18 08 39.33	&  -20 24 39.94	&  7.55 & 10* & 18 & 15 \\
 \sgrb & 19 07 14.33	 & +09 19 20.1	 & 5.17 & 100* & 6.5 & 15 \\
 \sgrc & 05 26 00.89	 & -66 04 36.3	&  8.05 & 65  & 7.3 & 55 \\
 \sgrd $^o$ & 16 35 51.84	 & -47 35 23.3	 & 2.59 & 19 & 2.2 & 11 \\
 \sgre $^o$ & 05 01 6.78	 & +45 16 34.0	 & 5.76 & 6.8 & 2.0 & 5.0 \\
 \sgrf $^o$ & 04 18 33.86	 & +57 32 22.91 &  9.08 &  $<$0.006 & $<$0.075 & 2.0 \\ 
 \sgrg $^o$ & 18 33 44.38	 & -08 31 07.71	 &  7.56 & 4.0 & 1.8 & 10 \\
\psr  &  16 22 44.8 &  -49 50 54.4  & 4.32  & 17 &  2.8 & 9.0 \\
\hbpsr  & 18 46 24.94 & -02 58 30.1 & 0.32 & 7.1 & 0.5 & 6.0 \\
  \axj$^{o,a}$ & 18 44 54.68	&  -02 56 53.1	 & 6.97 & -  & - & 8.5 \\
\noalign{\smallskip}\hline\noalign{\smallskip}
\end{tabular}
$^*$ Variable parameters; see http://www.physics.mcgill.ca/$\sim$pulsar/magnetar/main.html for some of the alternative values. \\
$^o$ Sources which showed outburst activity (discussed in \S\,5).  \\
$^a$ Candidate magnetar.\\
$^b$ We report here on the most recent values for the distances, however we caveat that for most of these objects the distance determination is very uncertain or controversial. We used these distances to infer the luminosity in Tab.~2, which can be easily scaled if in the future more precise distances will be available.
\end{table}
%


\section{Multi-band view of magnetars}

Until about 10 years ago, magnetars were thought to be emitting
exclusively in the X-ray energy range, while having sporadic flares
reaching the soft $\gamma$-ray energies. It was only recently that the
availability of new instruments, as well as a progressively better
understanding of these objects, prompted the search and the following discovery
of their emission at other wavelengths. Below we summarize the multi-band properties of magnetars.

\subsection{Radio emission of magnetars}

For a long time magnetar candidates were thought to be radio
quiet. However, their radio detections started about ten years ago, when radio observations of one SGR (\sgra )
performed with the Very Large array (VLA), revealed the presence of
elongated structures with a variable shape and orientation over a year
timescale \citep{vasisht94,frail97}. These jet-like features suggested that outflows from \sgra\,
have been occurring, most probably connected with its conspicuous
bursting and flaring activity. In fact, after a few years transient
radio emission were discovered from SGRs which showed giant flares
(see \citep{rea08,mereghetti08} and reference therein). Only three
these events were detected in the last 30 years and after two of them
a transient radio counterpart was discovered \citep{frail99,gaensler05}, while in the first case no data were
available. In particular, radio observations after the recent giant
flare from \sgra\, \citep{palmer05,hurley05} clearly
showed a radio structure moving out from the source and changing in
polarization, believed to be an outflow \citep{gelfand05,taylor05,fender06}.

Furthermore, recently pulsed radio emission were also discovered in a
few magnetars. The first detection of pulsed radio emission was from
the transient AXP, \xte , which has been first detected in the radio
band as a continuum source (4.5mJy; \citep{halpern05}), and soon
after recognized as a radio pulsar \citep{camilo06, ccr07}. Another transient AXP has been observed to show radio pulsed
emission, \aa\, \citep{camilo07}, with similar characteristics as
the \xte . In particular, in radio magnetars show many
properties at variance with canonical radio pulsars, as a flat
spectrum, large flux variability (until a factor of $\sim$10) on
timescales as short as fraction of hours, and a transient behaviour
thought to be connected with the X-ray outburst of the source.

However, the discovery of the third radio magnetar, \psr\, \citep{levin10}, has instead showed that the strong connection between the
transient pulsed radio emission and the X-ray outbursts of magnetar
candidates does not always necessarily hold. In fact \psr\, has shown
in the past 7 years all typical radio-magnetar emission, it has a field of $B\sim3\times10^{14}$\,Gauss, but its X-ray
counterpart was never detected in outburst thus far.

For almost all other magnetars deep upper limits were derived on their radio
pulsed emission \citep{burgay06,crawford08,eit10}.

\subsection{Optical and infrared emission of magnetars}

The availability of large telescopes as the Very Large Telescope,
Gemini or Keck, made it feasible the detection of the weak optical and
infrared counterparts to magnetar candidates (see \citep{mignani10} for a recent review). Furthermore, the new
adaptive optics technology allowed the detection of such counterparts
even in the very crowded regions of the Galactic plane.

The first discovery of such counterparts was ten years ago, with the detection of \uu\, in the optical band \citep{hulleman00}.

An exact modeling of the entire spectral energy distribution of a magnetar has not yet been
done. In the low energy bands, an infrared  ``excess" (or ``flattening") is seen with respect to the extrapolation of the blackbody often used to account for the optical emission. However,
instead if we connect the infrared points with the non-thermal hard X-ray
observations (as for the typical radio pulsars), we end up with an excess in the optical band. Thus, whether this optical and infrared emission comes from a
thermal process or not is still unclear and would be an important
information for constraining the models: in fact, if this emission
comes from a disk it is expect to be thermal \citep{perna00}, while if it comes from processes on the magnetosphere
of a magnetars we expect the optical spectral continuum to be non
thermal \citep{ertan04}. 

%
\begin{table}[t]
\caption{X-ray spectral properties and luminosities in different wavebands.}
\label{tab:2}       
%
%
\begin{tabular}{p{2.8cm}p{1.2cm}p{0.8cm}p{0.8cm}p{1.1cm}p{1.1cm}p{1.0cm}p{1.0cm}p{1.7cm}}
\hline\noalign{\smallskip}
Magnetar & $kT_1$/$kT_2$  & $\Gamma_1$  & $\Gamma_2$ & $L_{\rm 35, soft}^b$ & $L_{\rm 35, hard}^b$& $S~d^2$ & $K_s$ & References \\
 &  (keV) &  & & (erg s$^{-1}$) & (erg s$^{-1}$) & (mJy kpc$^2$) & mag\\
\noalign{\smallskip}\svhline\noalign{\smallskip}
\ea $^o$   &  0.4*& 4.1* & -- & 0.2* & -- & -- & 21.7* & \citep{woods04,tam04}\\
 \uu  $^o$ & 0.4* & 3.9* & 0.9*& 3* & 0.6*  & -- & 19.7 & \citep{rni07,denhartog08,hulleman04} \\
 \rxs & 0.5* & 3* & 1.1* & 1.4* &0.7 & -- & -- & \citep{rea03,roz05,gri07} \\
 \ee $^o$ & 0.6* & 3.4* & -- & 0.08* & -- & -- & 19.4* & \citep{tmt05,wang02} \\
 \kes & 0.4	 & 2 & 1.5 & 2.6 & 2.6 & -- & -- & \citep{morii03,gotz06} \\
 \cxo & 0.3/0.7	 & -- & -- & 1.5 & -- & -- &  -- & \citep{tiengo08} \\
 \wes $^o$ & 0.6* & 2.3* & -- & 2.5* & -- & -- & -- & \citep{icd07}\\ 
 \xte $^o$ & 0.3/0.7* & --& --& 0.35* & -- & 80* & 20.8* & \citep{gotthelf05,camilo06,irm04} \\
 \aa $^o$ & 0.6*& 1.7* & 1.5*& 1.6* & 2.4* & 40* & 18.5* &  \citep{camilo07,irr09,enoto10}\\
  \hess & 0.38 & 3.4 & -- & 0.2 & -- & -- & -- & \citep{hg10}\\
 \sgra $^o$ & 0.8 *& 1.2*& 2 & 12* & 12* & -- & 19.3* & \citep{met07,israel05} \\
 \sgrb  & 0.5* & 1.9* & 3.1 & 2.3* & 4 & -- & -- & \citep{met06,gotz06}\\
 \sgrc & -- & 3.3	& -- & 4 & -- & -- & -- & \citep{tiengo09} \\
 \sgrd $^o$ & 0.5* & 0.6* & -- & 0.15* & -- & -- & -- & \citep{esposito09} \\
 \sgre $^o$ & 0.7* & 2.9* & 0.8 & 1.8* & 1.1* & -- & 19.1 & \citep{rea09,rol08} \\
 \sgrf $^o$ & 0.9* & 2.6* & -- & 0.042* & -- & -- & -- & \citep{esposito10,eit10,rea10}\\ 
 \sgrg $^o$ & 1.2 & -- & -- & 0.9* & -- & -- & -- & \citep{eit10} \\
\psr  & 0.3 & -- & --  & 0.018 & -- & 390* & -- & \citep{levin10} \\
\hbpsr  & 0.9* & 1.9* & 2$^c$ & 3.1* & 2.8$^c$ & -- & -- & \citep{ng08,leahy08,mcbride08} \\
  \axj$^{o,a}$ & 0.6 & -- & -- & 1* & -- & -- & -- & \citep{gotthelf98} \\
\noalign{\smallskip}\hline\noalign{\smallskip}
\end{tabular}
$^*$ Variable parameters; see http://www.physics.mcgill.ca/$\sim$pulsar/magnetar/main.html for some of the alternative values. \\
$^o$ Sources which showed outburst activity (discussed in \S\,5).  \\
$^a$ Candidate magnetar.\\
$^b$ The soft and hard X-ray luminosities are given in the 1--10 keV  and 20--100 keV ranges, and in units of $10^{35}$ \ergs . The radio flux $S$ is calculated at 1.4 GHz. If only absorbed flux was found in literature or the flux was given for ranges other than 2--10 or 20--100 keV, XSPEC was used to estimate the unabsorbed flux.  Luminosities assume the distances given in Table~\ref{tab:1}.\\
$^c$ Including the (dominant) contribution from the pulsar wind nebula, that cannot be resolved from the pulsar emission by the current-generation gamma-ray instruments.
\end{table}
%


\subsection{Soft X-ray emission of magnetars}

Magnetar X-ray emission may be qualitatively separated into two components, a low-energy, $<$10 keV, and a higher energy
one, $>$20 keV. It is likely, although not proved yet, that different emission mechanisms are responsible for the two components. The low energy component is typically fit with either a blackbody with a temperature
kT $\sim$0.3--0.6 keV and a power-law with a relatively steep photon index, $\Gamma\sim$ 2--4, or two blackbodies with
kT$_1\sim$ 0.3 keV and kT$_2 \sim$ 0.7 keV \citep{mereghetti08}. In a few cases the low-energy component of SGR spectra has been fit with a single power-law, but recent longer observations have shown that, also for these sources, that an additional  blackbody component is required \citep{mte05}.

Thompson, Lyutikov \& Kulkarni \citep{thompson02} first pointed out that resonant scattering in magnetar magnetospheres may explain the
non-thermal emission observed in magnetar candidates. Due to the presence of hot plasma in the neutron star coronae, the thermal emission from the neutron star surface/atmosphere gets distorted through efficient resonant cyclotron scattering. 

Recent applications of resonant scattering models to the soft X-ray emission of magnetars have shown that indeed this interpretation fits well the data \citep{rea08,guver08,zane09}, and finds that these sources are characterized by magnetospheric plasma with a density which, at resonant radius, is about 3 orders of magnitude higher than the Goldreich-Julian electron density.

\begin{figure*}[t]
\includegraphics[width=11cm]{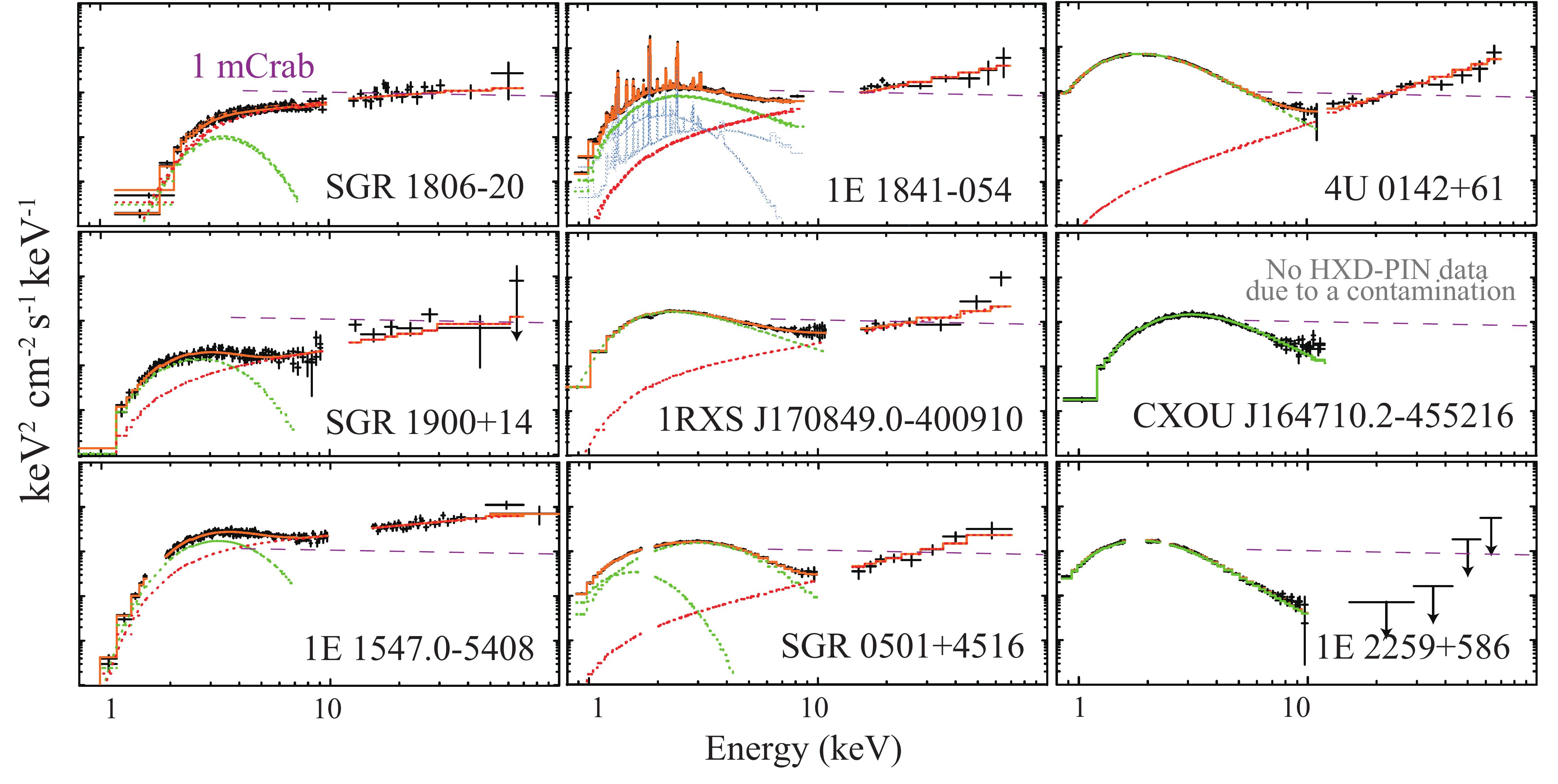}
\caption{Hard X-ray emission of magnetars as observed by Suzaku (from \citep{enm10}). }
\label{hardX}     
\end{figure*}


\subsection{Hard X-ray emission of magnetars}

Hard X-ray emission until $\sim$200 keV has been discovered for some magnetars (namely \ea , \rxs , \kes , \sgra ,\break \sgrb , \aa , \sgre\ \citep{kuiper04, kuiper06, mte05, denhartog08,gotz06,rea09,enoto10}; see also Figure \ref{hardX}), thanks to INTEGRAL and RXTE.
Hard X-ray variability has also been discovered in two cases: for \sgra, and recently in the \sgre\ (see \citep{gotz07,rea09}). In the two brightest source, a strong variability of the hard X-ray emission with the pulsar phase has been discovered \citep{denhartog08,dkh08}.

The discovery of magnetarsÕ hard X-ray emission opened a new window in the study of magnetars, making it crucial to revise the current knowledge of magnetars being steady soft X-ray sources. However, despite recent theoretical works \citep{beloborodov07,baring07}, a clear physical idea on how this hard X-ray emission is generated is still missing. Very promising are recent models invoking resonant cyclotron scattering in the relativistic regime \citep{nobili08,baring07} although current codes does not allow any observational tests so far.

\section{Magnetars' Outbursts}

In this section we report one by one on the outbursts that have been detected so far from magnetars. We define here as outbursts an increase in the persistent level of at least a factor of 5. However, note that many of the sources not listed below showed smaller outbursts, often connected with bursting activity (as e.g. \kes, \hbpsr, and \sgrb; \citep{ks10,gavriil08,ks08,esposito07,ggo11}). A comprehensive image of the decay of most these outbursts (the ones successfully followed by X-ray imaging instruments) are reported in Figure \ref{outbursts}.

\subsection{\ea}

This 7\,s X-ray pulsar was discovered a few decades ago, and it was believed to be a very stable X-ray emitter and pulsator. However, in 2002, it emitted the first notable recorded case of flux variability from a magnetar \citep{kaspi03}. Its active phase, with a factor of $\sim$10 persistent flux enhancement, was followed by the onset of a bursting activity phase during which the source displayed more than 80 short bursts \citep{gavriil04,woods04} . Furthermore, despite being the first case of an outburst from an otherwise persistent and quiet magnetar candidate, \ea\, was also the first to show a connected X-ray and infrared outburst \citep{tam04}. It showed a glitch during the enhanced activity, corroborating the idea of these outbursts being due to crustal stresses imparted by the unstable magnetic fields \citep{kaspi03}. Furthermore, this magnetar lies in the supernova remnant (SNR) G109.1-1.0 (CTB 109), being one of the few associations between magnetars and SNR which still holds \citep{gaensler01} .

\subsection{\uu}

4U 0142+614 is one of the brightest magnetar known to date, and it
was first detected by Uhuru in 1978. However,
mainly because of the presence of the accretion-powered binary
pulsar RX J0146.9+6121 nearby, only in 1994 was an $\sim$8.7 s periodicity
reported using EXOSAT data taken in 1984 \citep{israel94}.
Long-term spin-period variations were discovered
thanks to a large RXTE campaign \citep{gavriil02}, leading
to the measure of the period derivative $\dot{P}\sim 2 \times 10^{-12}$ \ss .
Despite deep searches \citep{israel94,wilson99}, no evidence
for orbital motion has been found, supporting the isolated neutron star
scenario. Further observations \citep{white87,israel99,paul00} revealed a soft X-ray spectrum typical
of an AXP, best fitted by an absorbed blackbody (kT$\sim$0.4 keV) plus a power-law ($\Gamma\sim$3.7). More recent Chandra \citep{juett02,patel03}, \xmm\ \citep{gohler05,rni07} and Swift X-ray Telescope (XRT) \citep{rea07} observations
have shown that \uu\, is a relatively stable X-ray emitter (although recently it showed an outburst). In the last few years, two peculiar characteristics of
\uu\, have been found in comparison with other AXPs: (i)
an optical counterpart \citep{hulleman00} displaying 8.7 s pulsation with a 30 per cent pulsed fraction \citep{kern02}
 and (ii) mid-infrared emission, tentatively interpreted
as the signature of a non-accreting disc around the neutron star \citep{wang06}. Furthermore, like in other magnetars, a
hard X-ray emission up to 250 keV has been revealed \citep{denhartog06,denhartog07}.

\subsection{\ee}
The 6.4\,s X-ray pulsar \ee\, was serendipitously discovered by Einstein during observations of the Carina Nebula \citep{seward86}. Immediately suggested to be highly variable,\footnote{The flux variability of \ee\, had been long debated and could be confirmed only twenty years later \citep{mereghetti04,gavriil04}.} it was tentatively classified as a binary with a $V\sim19$ mag Be companion \citep{seward86}. Subsequent observations however ruled out the candidate optical counterpart and \ee\,  was placed in the emerging (at the time) class of the anomalous X-ray pulsars \citep{hellier94,vanparadijs95,mereghetti95}.  During October/November 2001 a couple of SGR-like bursts -- the firsts ever found in an AXP -- were detected from \ee\, with RXTE, virtually unifying the SGR and AXP classes \citep{gavriil02}. 

1E\,1048.1--5937 is one of the most frequently observed magnetars and in the last decade in particular it has been extensively monitored with RXTE since, because of its unstable spin down, frequent RXTE observations are necessary in following the spin evolution \citep{gavriil04}. The long-term light curve of \ee\,  shows two consecutive outbursts in the 2002--2004,  and another event in 2007 \citep{tam08}.

 The late 2001 bursts marked the start of the first stretch of enhanced flux that persisted a few months \citep{gavriil04}. The peak flux was $\sim$2 times the mean quiescent value ($\approx$$7\times10^{-12}$  erg cm$^{-2}$ s$^{-1}$). Another larger (the peak flux reached $\sim$3 times the quiescent one) and longer-lived flux increase started in Spring 2002 and lasted into 2004. A third burst was observed from \ee\, in June 2004, during the final phases of the outburst. Both events had a few-weeks-long rise time and much longer and gradual decays. These flux variations were accompanied by substantial timing irregularities (including glitches) which, however, do not correlate in an obvious way with the flux enhancements \citep{gavriil04}. On the other hand, anti-correlation between pulsed fraction and flux and a correlation between spectral hardness and flux have been reported for this source \citep{tmt05,tam08}.
 
  In March 2007 \ee\, entered a new outburst accompanied by a large spin-up glitch \citep{dib09}. This time, the flux rose to the peak (slightly higher than that of the 2002--2004 outburst) in less than a week. Also a fourth burst was detected by RXTE about a month after the outburst onset \citep{dib09}. For a distance of 9 kpc \citep{durant06}, the total energy emitted has been estimated in roughly $4.8\times10^{41}$, $3.5\times10^{42}$, and $4.3\times10^{42}$ erg for the three outbursts in chronological order \citep{dib09}. We note that Gaensler et al.~\citep{gmo05} propose an association between \ee\, and the hydrogen shell GSH 288.3--0.5--28, which would pose the source at a much shorter distance of 2.7 kpc.

A candidate infrared counterpart to \ee\, ($K_{\rm{s}}=19.4$ mag) was selected in 2001 with the Baade (Magellan I) telescope\citep{wang02} and confirmed by subsequent observations of large variability ($\sim$2 mag) \citep{ics02,durant05}. Changes in the infrared flux were initially suggested to be anti-correlated with those in the X-ray flux \citep{durant05}. Recent observations however showed a behavior inconsistent with this hypothesis, with infrared flux enhancements near those at X-rays, indicating that neither disposition holds all the time \citep{tam08,wang08}.


\begin{figure*}[t]
\includegraphics[width=10cm,angle=-90]{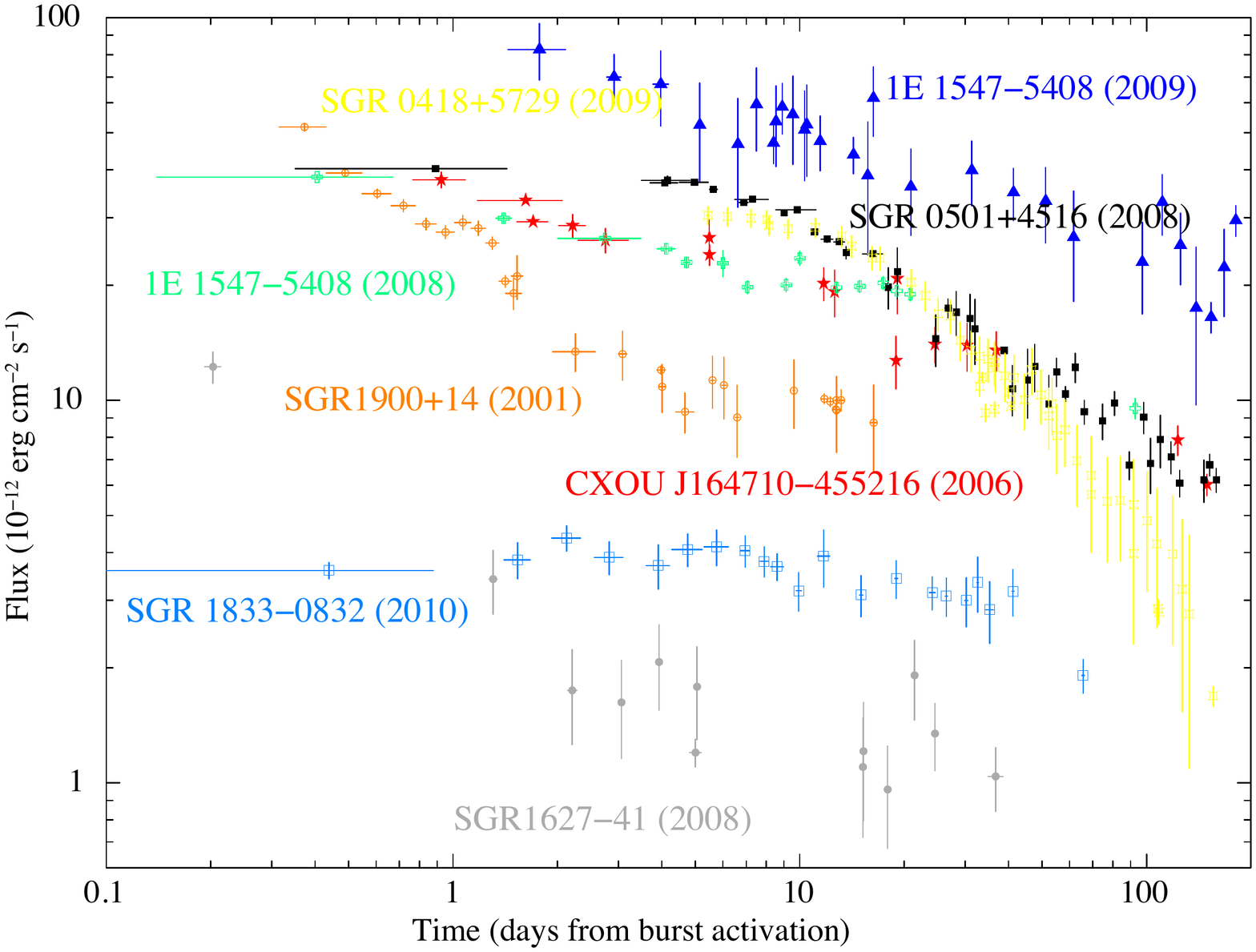}
\caption{Flux evolution over the first $\sim200$\,days of recent
  magnetar outbursts (all observed with imaging instruments). Fluxes
  are reported in the 1--10 keV energy range, and the reported times
  are calculated in days from the detection of the first burst in each
  source. In particular we show \wes\, in red \citep{israel07},
  \sgrd\, in grey \citep{esposito08}, \sgra\, in orange \citep{feroci01}, \aa\, in green and blue for the 2008 and 2009 outbursts,
  respectively \citep{israel10,ng10},
  \sgre\, in black \citep{rea09}, \sgrf\, in yellow \citep{esposito09}, and \sgrg\, in light blue \citep{gogus10,eit10}.}
\label{outbursts}     
\end{figure*}


\subsection{\wes}

\wes\, was discovered during deep X-ray observations of the massive star cluster Westerlund 1 \citep{muno06}. In 2006 it underwent a large outburst \citep{muno06,muno07}, preceded by a bright X-ray burst detected with the Swift observatory on 2006 September 21. This  10.6 s X-ray pulsar is located in the young cluster of massive stars, Westerlund 1 \citep{muno06}. Several observations with multiple X-ray telescopes were performed
following the burst detection, showing a the flux from this magnetar increased by a factor of $\sim$100 following the outburst. \xmm\, observed the source just 4 days prior to the outburst, providing one of the few cases where the beginning of the outburst could be assessed with relative precision. Muno et al. \citep{muno07}  showed that the spectrum of \wes\, 
hardened significantly when the flux increased and that the pulse profile changed dramatically
from a simple near-sinusoidal shape to a complex profile with three distinct peaks per cycle. A glitch as large as $\Delta\nu / \nu > 1.5\times10^{-5}$ was detected in coincidence with the outburst (\citep{israel07} but see also \citep{woods10}).

\subsection{\xte}

It was only in 2003 that the first transient magnetar was discovered, namely\break XTE\,J1810--197, which displayed a factor of $>$100 persistent flux enhancement with respect to the unpulsed pre-outburst quiescent luminosity level ($>10^{33}$\ergs ; \citep{ibrahim04,israel04,gotthelf04}).
Unfortunately, the initial phases of the outburst were missed and we do not know whether a bursting activity phase occurred in coincidence with also for this source. During the latest 5 years it was extensively observed, covering a flux variability over a factor of about $>$60. Since the very first 2003 observations of XTE\,J1810--197, carried out approximately one year after
the onset of the outburst, it was evident \citep{gotthelf04} that the source spectral shape
(initially of two blackbodies with kT = $0.29\pm0.03$ keV and kT = $0.70\pm0.02$ keV) was significantly different from that serendipitously recorded by ROSAT in 1992 (one blackbody with kT $\sim$ 160 eV \citep{gotthelf04}). Moreover, the source showed a 5.54 s pulsation with a pulsed fraction of nearly 45\% during outburst, while an upper limit of 24\% was inferred from
the ROSAT data. 

In 2006 the source was discovered to be one of the most intense and
polarized radio pulsar in our sky with single peak flux density reaching a few Jy \citep{camilo06}.
This finding provided direct evidence that the radio pulsar emission can also be at
work in magnetars, corroborating the analogy with the rotational powered pulsars. Furthermore, it strongly suggested that a better way to study these objects (when in a high state)
is taking into account for the whole emission properties of the source from radio to the hard X-ray.
Simultaneous X-ray and radio observations have been performed. These suggested that the X-ray and radio emitting
regions are likely coincident (or superimposed), the X-rays likely coming from a larger area. Moreover,
during all these campaigns large radio flux ($\sim$50\%) and pulse shape variations have been detected which
do not correlate with any change (at a few percent level) of the X-ray timing and/or spectral parameters (Israel et al. in preparation).
This suggests that the X-ray emission likely originates deep in the crust (or more in general, the radio and
X-ray mechanisms are not closely related).

\begin{figure*}[t]
\includegraphics[width=11cm,height=7cm]{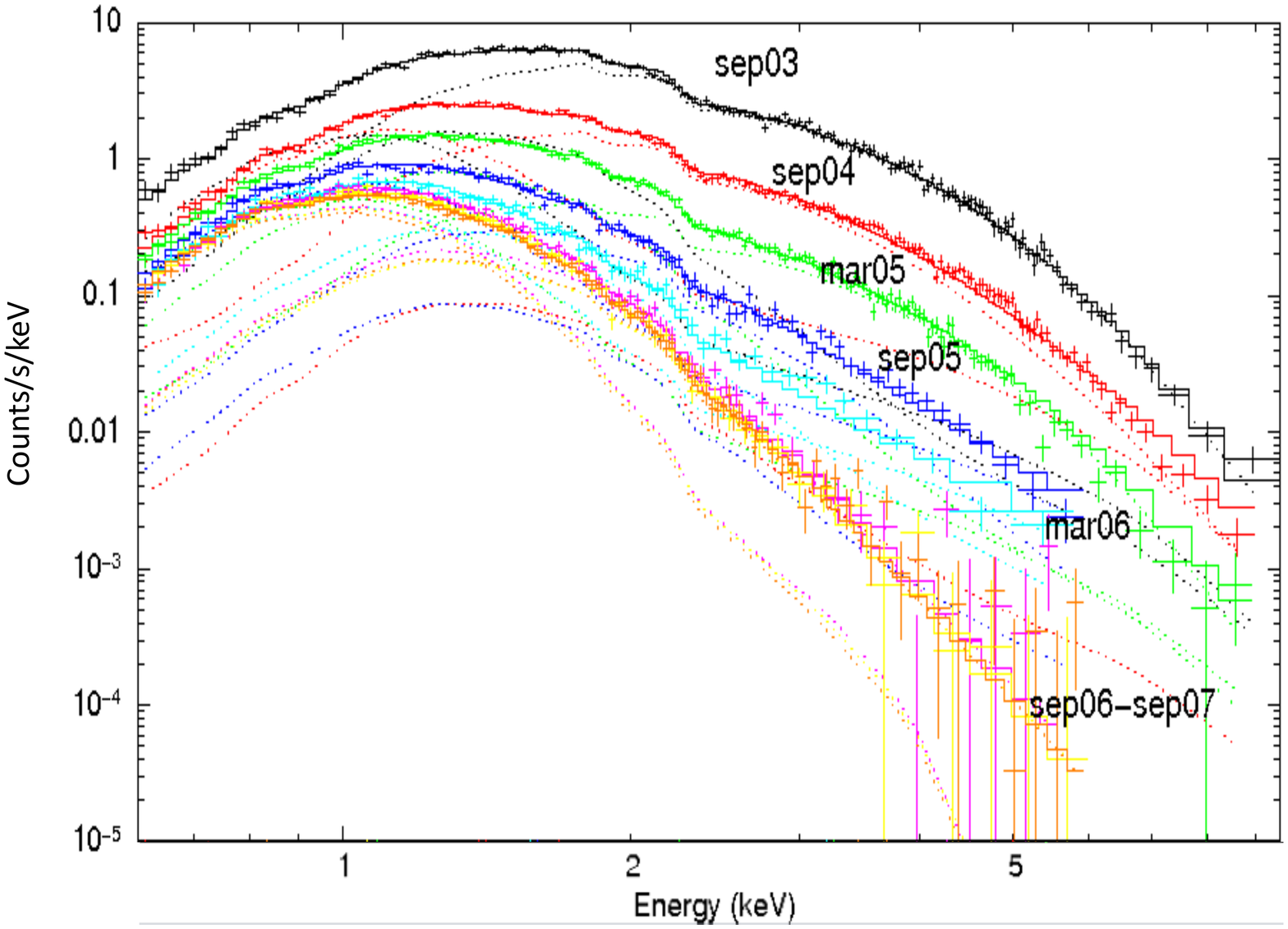}
\caption{Spectral evolution of \xte\, over several years \citep{albano10}.}
\label{spectra}     
\end{figure*}

A variable infrared counterpart have been discovered for this object \citep{israel04,rea04}, although no clear correlation between the X-ray and infrared variability have been confirmed \citep{testa08}. \xte\, is now close to its quiescent level, and showed the longest outburst decay ever observed in magnetars, with a consequent slow spectral softening over a timescale of about 5 years (see also Figure\,\ref{spectra}).

\subsection{\aa}

\aa\, was first proposed as a possible magnetar in the candidate supernova remnant G327.24--0.13 through X-ray observations \citep{gelfand07} and subsequently recognized as a transient radio magnetar thanks to the discovery at radio frequencies of its spin period of  $\sim$2.1 s \citep{camilo07} (later detected also in X-rays \citep{halpern08}).
In recent years, 1E\,1547--5408 has been one of the most active magnetars. A first outburst occurred during the Summer 2007, when Swift observations caught 1E\,1547--5408 at an X-ray flux level of $\sim$$5\times10^{-12}$ erg cm$^{-2}$ s$^{-1}$, more than one order of magnitude brighter than in quiescence \citep{halpern08}. However, the early phases of this outburst were missed and no bursts were observed (possibly due to a sparse X-ray coverage).

A new outburst started on 2008 October 3 \citep{israel10,ng10}. This time several bursts were detected by Swift and in the data taken immediately after the Swift trigger, 1E\,1547--5408 was found at a flux level of $\sim$$6\times10^{-11}$ erg cm$^{-2}$ s$^{-1}$. Then the luminosity declined by 70\% in three weeks, following a power-law fading trend with decay index $\sim$$-0.2$ \citep{israel10}. During this period, the source displayed a complex timing and spectral variability \citep{israel10,ng10}.

No further bursts were reported until 2009 January 22, when the source entered a new stretch of much stronger activity, with thousands of bright bursts detected by many instruments \citep{mereghetti09,kaneko10,savchenko10,ng10}. A spectacular event connected with this giant outburst was the appearance around the source of multiple expanding X-ray rings due to scattering by different layers of interstellar dust of a particularly bright burst \citep{tiengo10}. From the analysis of these structures, a distance to the source of $\sim$4 kpc was proposed \citep{tiengo10}.\footnote{This value agrees with the distance suggested by the possible association of 1E\,1547--5408 with the supernova remnant G327.24--0.13 \citep{gelfand07}.} Pulsations were detected up to $\sim$150 keV \citep{kdh09,kaneko10}, and the source reached a flux level of $\sim$$8\times10^{-11}$ erg cm$^{-2}$ s$^{-1}$ (see Figure~\ref{outbursts}). The (ongoing) intensive X-ray monitoring of 1E\,1547--5408 shows that the flux is slowly decaying with an overall power-law trend with index $\sim$$-0.3$.

As periods of outburst activity are the most promising to search for radio and optical/infrared emission from magnetars, the 2009 outburst of 1E\,1547--5408 triggered several multi-wavelength follow ups. Multiple radio observations were carried out at Parkes on 2009 January 22, 23 and 25. Pulsed emission from 1E\,1547--5408 was detected at 3 GHz during a 1.2-hour-long observation on January 25, but, notably, not in the other two occasions \citep{burgay09}. Moreover, a relatively bright transient (near) infrared source ($K_{\rm{s}}\sim18.5$ mag) was discovered with ESO/VLT within the radio positional uncertainty of the AXP and identified with its counterpart \citep{irr09}. Deep infrared observations taken with ESO/VLT during the 2007 outburst
revealed four objects consistent with the radio position of 1E\,1547--5408 \citep{mignani09}; none of them, however, showed variability and the likely counterpart was not detectable at the time. This sets an upper limit of about 21 mag in the $K_{\rm{s}}$ band on the infrared emission of the AXP during the 2007 outburst.


\subsection{\sgra}

\sgra\, is arguably the most burst-prolific magnetar and showed several periods of bursting activity since the time of its discovery in 1979
 \citep{laros86,laros87}. Its persistent X-ray counterpart was observed for the first time with the ASCA satellite in 1993 \citep{murakami94}. Subsequent observations with RXTE led to the discovery of coherent pulsations (the first time in an SGR) at $P\simeq7.5$ s and a secular increase of the period at a rate of $\sim$$8\times10^{-11}$ s  s$^{-1}$ \citep{kouveliotou98}. These values confirmed the debated neutron star nature of the SGRs and, since under the assumption of pure magnetic dipole braking they imply a surface magnetic field strength of \mbox{$8\times10^{14}$ G}, provided strong support for the magnetar model, that was developed in the early 1990s \citep{duncan92,paczynski92}.\\
\indent The source luminosity remained fairly constant for many years at $\sim$$5\times10^{35}$ erg s$^{-1}$ (for a distance of 15 kpc \citep{corbel97,mcclure05}), until both the burst rate and the X-ray persistent emission started increasing during 2003 and throughout 2004, when the luminosity approximately doubled with respect to the ``historical'' level \citep{mte05,woods07}. This period of intense activity culminated with a giant flare recorded on 2004 December 27 \citep{hurley05,mereghetti05,palmer05}.
This giant flare was exceptionally intense (assuming isotropic luminosity $\sim$$10^{47}$ erg were released) and produced strong disturbances in the Earth's ionosphere \citep{campbell05,inan07} and detectable effects on the geomagnetic field \citep{mandea06}.\\
\indent The initial flash was followed by a tail clearly modulated at the spin frequency of SGR\,1806--20 that persisted for $\sim$380 s. Comparing this giant flare with those seen from SGR\,0526--66 and SGR\,1900+14, it is found that the energy in the pulsating tails of the three events was roughly of the same order ($\sim$$10^{44}$ erg), while the energy in the initial spike of SGR\,1806--20 (a few $10^{46}$ erg) was at least two orders of magnitude higher than that of the other events.\\
\indent Observations with RXTE unveiled, for the first time in an isolated neutron star, rapid quasi-periodic oscillations in the pulsating tail of the flare, likely related to global seismic oscillations on the neutron star surface \citep{ibs05}. The flare was accompanied by the emission of relativistic particles which powered a synchrotron nebula (a ``mini-plerion") that faded in a few months \citep{cameron05,gaensler05,taylor05,fender06}. A similar ``radio afterglow" was observed also following the giant flare from SGR\,1900+14 \citep{frail99} (in the case of the 1979 event from SGR\,0526--66, no data at radio wavelengths were available).\\
\indent The small positional uncertainty of the radio observations permitted the identification of the infrared counterpart of the SGR \citep{kosugi05,israel05}. The fluxes observed in the infrared and gamma energy bands show a variability correlated with that observed in the 2--10 keV energy range \citep{met07}. After the giant flare, the persistent X-ray flux of SGR\,1806--20\ started to decrease from its outburst level, and its X-ray spectrum to soften, but the source has remained moderately burst-active to date \citep{rea05,met07,tiengo05,woods07,emt07}. During the Spring 2006 the source recovered its pre-outburst luminosity \citep{met07} and about five years after the flare (on 2009 September 7--8) the value measured with XMM-Newton was $\sim$$3\times10^{35}$ erg s$^{-1}$. A flux decrease have been observed also from its infrared counterpart \citep{israel05,rea05_atel,met07}.

\subsection{\sgrd}
SGR\,1627--41\ was discovered in 1998, when about one hundred bursts in six weeks were observed by CGRO/BATSE and other high-energy instruments \citep{woods99}. Soon after the discovery of the bursts, its soft X-ray counterpart was identified with BeppoSAX at  a flux level of $\sim$ $7\times10^{-12}$ erg cm$^{-2}$ s$^{-1}$\ (unabsorbed, 2--10 keV), corresponding to a luminosity of $\sim$ $10^{35}$ erg s$^{-1}$\ for a distance to the source of 11 kpc \citep{corbel99}. 
In the following 10 years no further bursting activity was reported while various observations carried out with BeppoSAX, ASCA, Chandra, and XMM-Newton\ showed a spectral softening and a monotonic decrease in the luminosity, down to a level of $\sim$$10^{33}$ erg s$^{-1}$\citep{kouveliotou03,mereghetti06,eiz08}.
 
The long-term fading of SGR\,1627--41 was suddenly interrupted by its burst reactivation on 2008 May 28, when several bursts were detected by the Swift/BAT during a short period of activity (about one day) \citep{eiz08}. This episode was associated with an abrupt and temporary large enhancement of the persistent X-ray flux (a factor of about 100 above the last measurement, in February 2008 with XMM-Newton) and a marked spectral hardening \citep{eiz08}. 
In September 2008, a deep XMM-Newton observation yielded the first measure of the spin period of the source (2.6 s, making SGR\,1627--41 the second fastest spinning magnetar after 1E\,1547--5408) and, together with searches in Chandra archival data, of its spin-down rate ($1.9\times10^{-11}$ s s$^{-1}$) \citep{esposito09,ebp09}. In fact, while the detection of strong SGR-like bursts from SGR\,1627--41\ made it a bona fide member of the SGR class, these two strong pieces of evidence in favor of the identification were still missing.\\
\indent Follow-up observations at near infrared and radio wavelengths were carried out in response to the 2008 burst activation, but they failed to detect the source \citep{deugarte09,camilo08atel1558}. In particular, the limit on the radio pulsed emission of SGR\,1627--41 at 1.4 GHz obtained at Parkes on 2008 May 30 and June 1 was 0.5 mJy (for a sinusoidal pulse profile) \citep{camilo08atel1558}.


\begin{figure*}
\vbox{
\includegraphics[height=2.5cm]{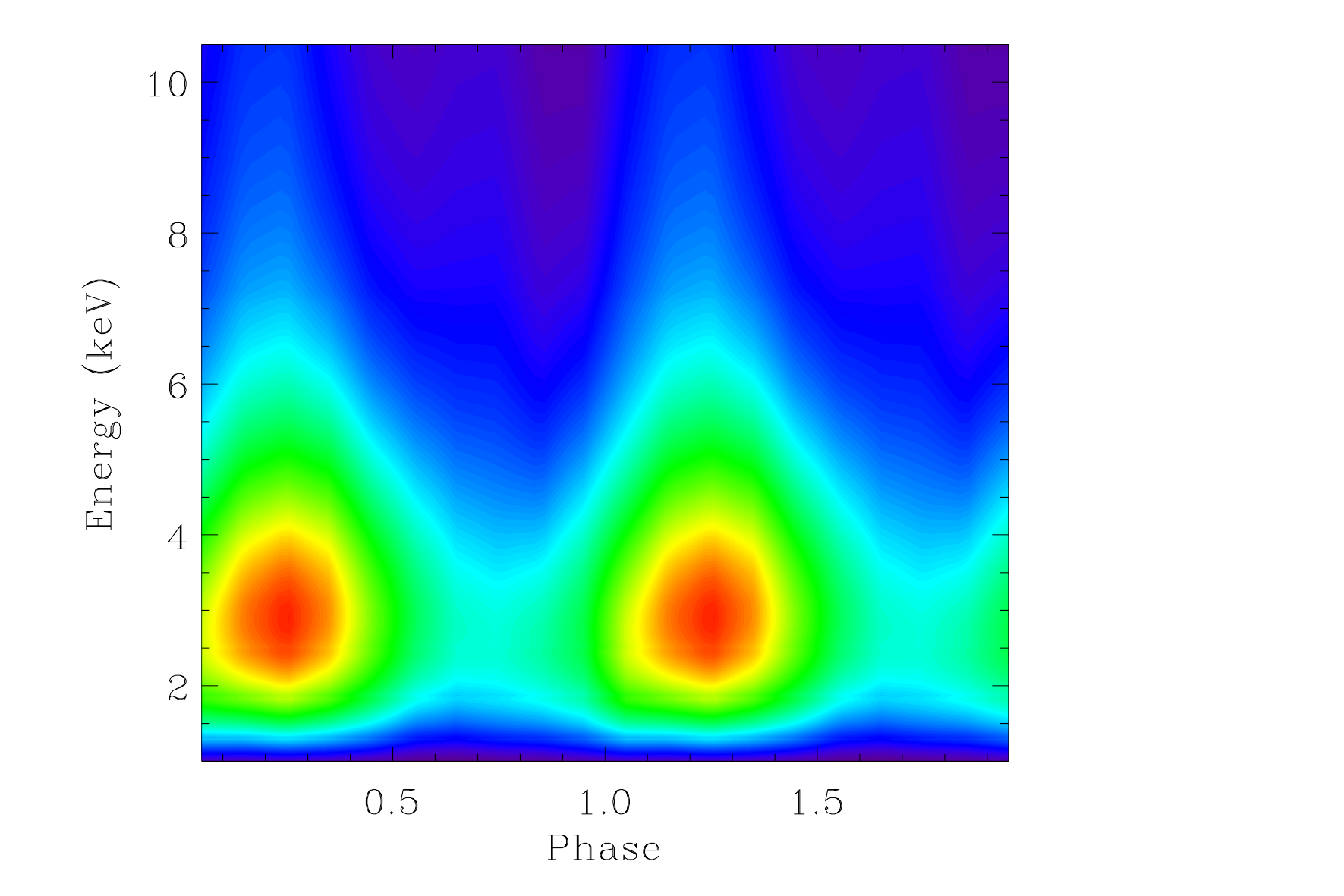}
\includegraphics[height=2.5cm]{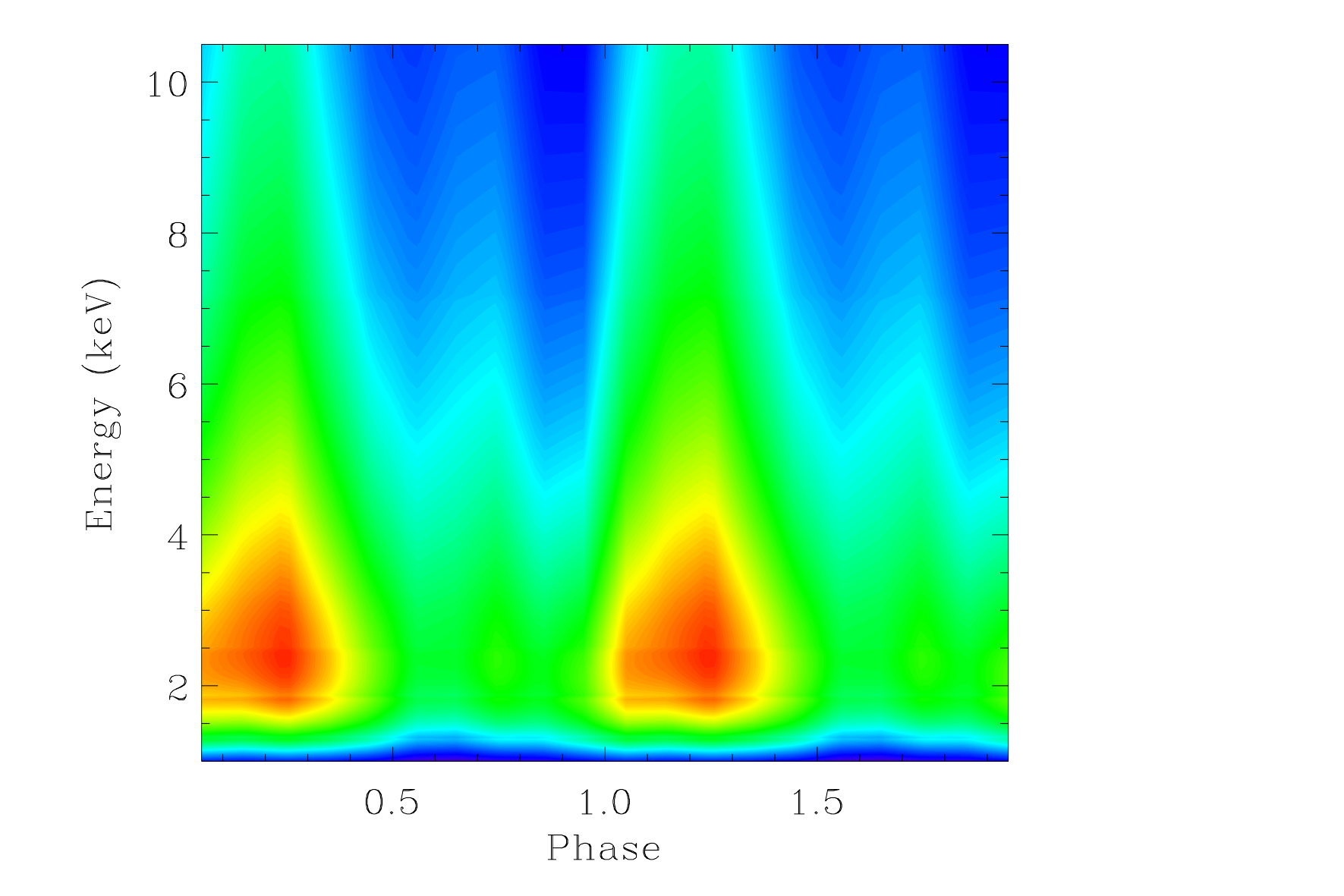}
\includegraphics[height=2.5cm]{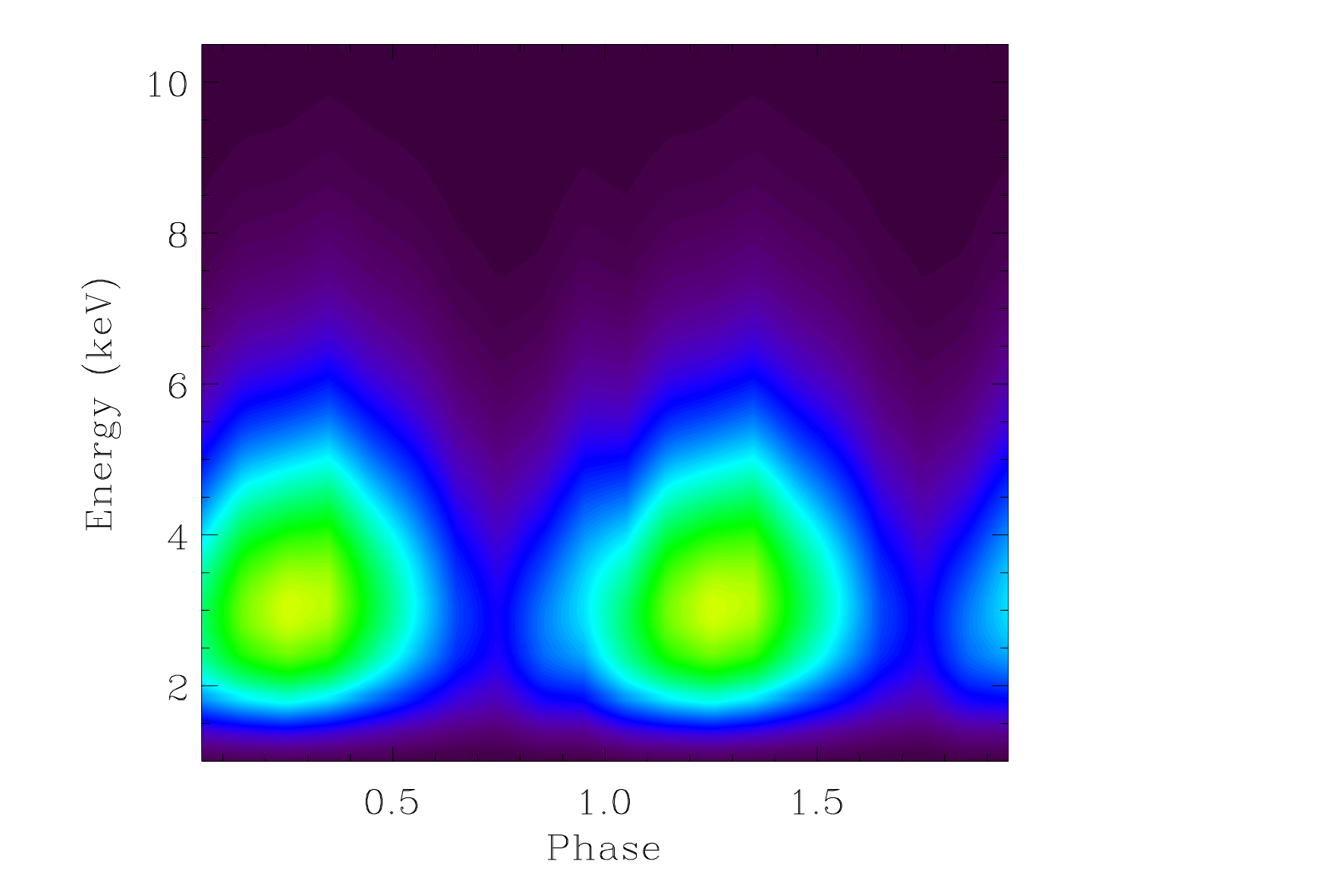}}
\vbox{
\includegraphics[height=2.5cm]{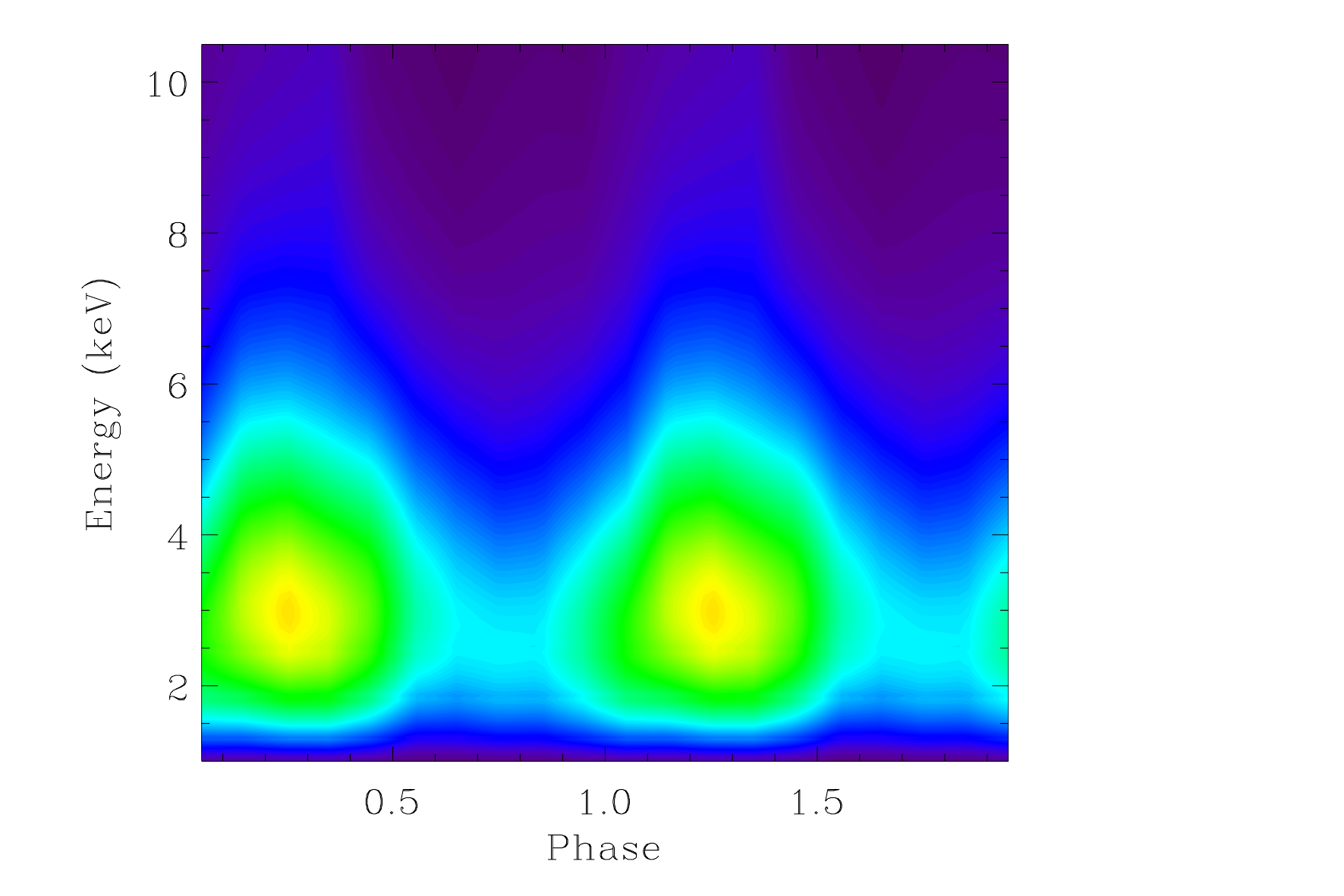}
\includegraphics[height=2.5cm]{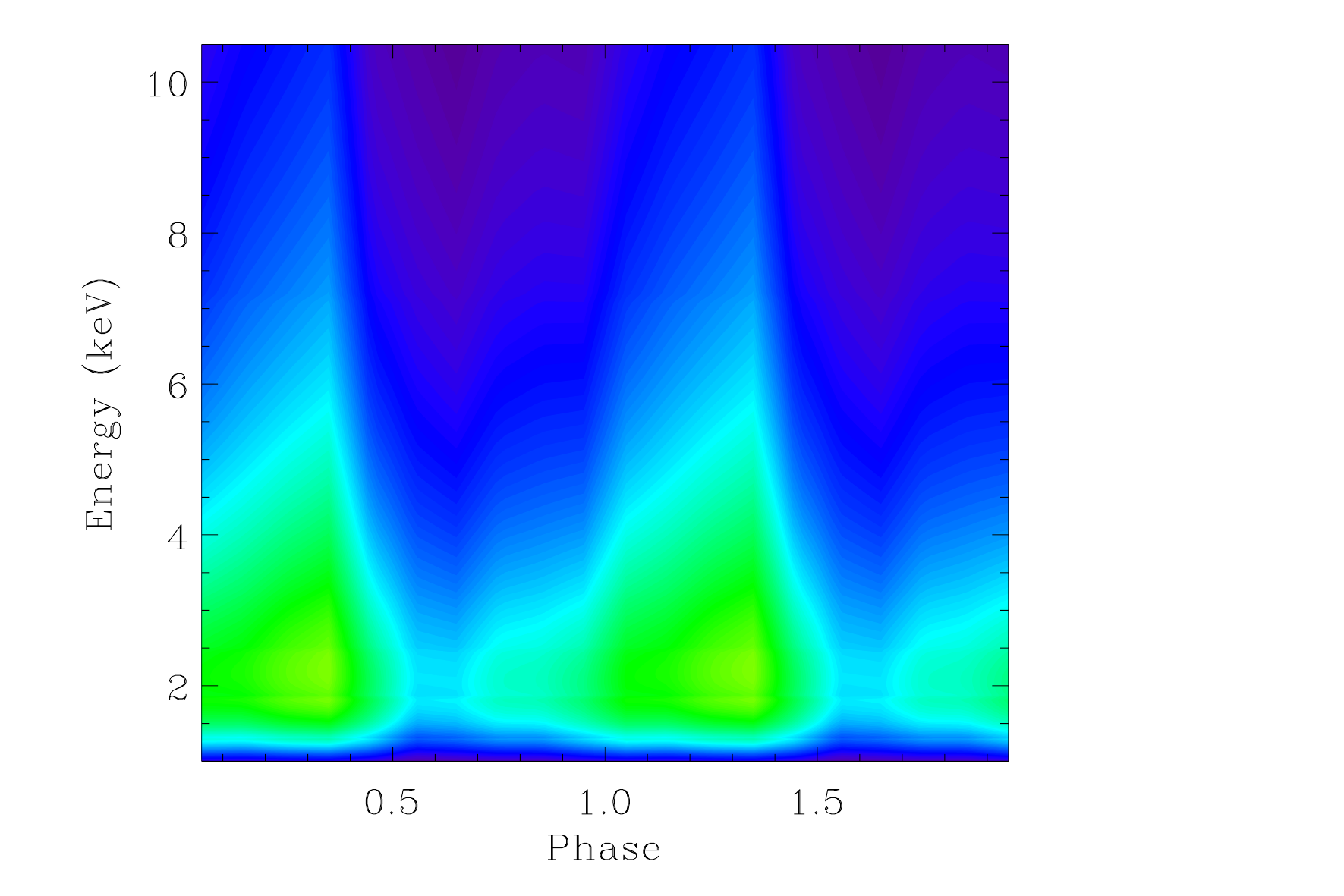}
\includegraphics[height=2.5cm]{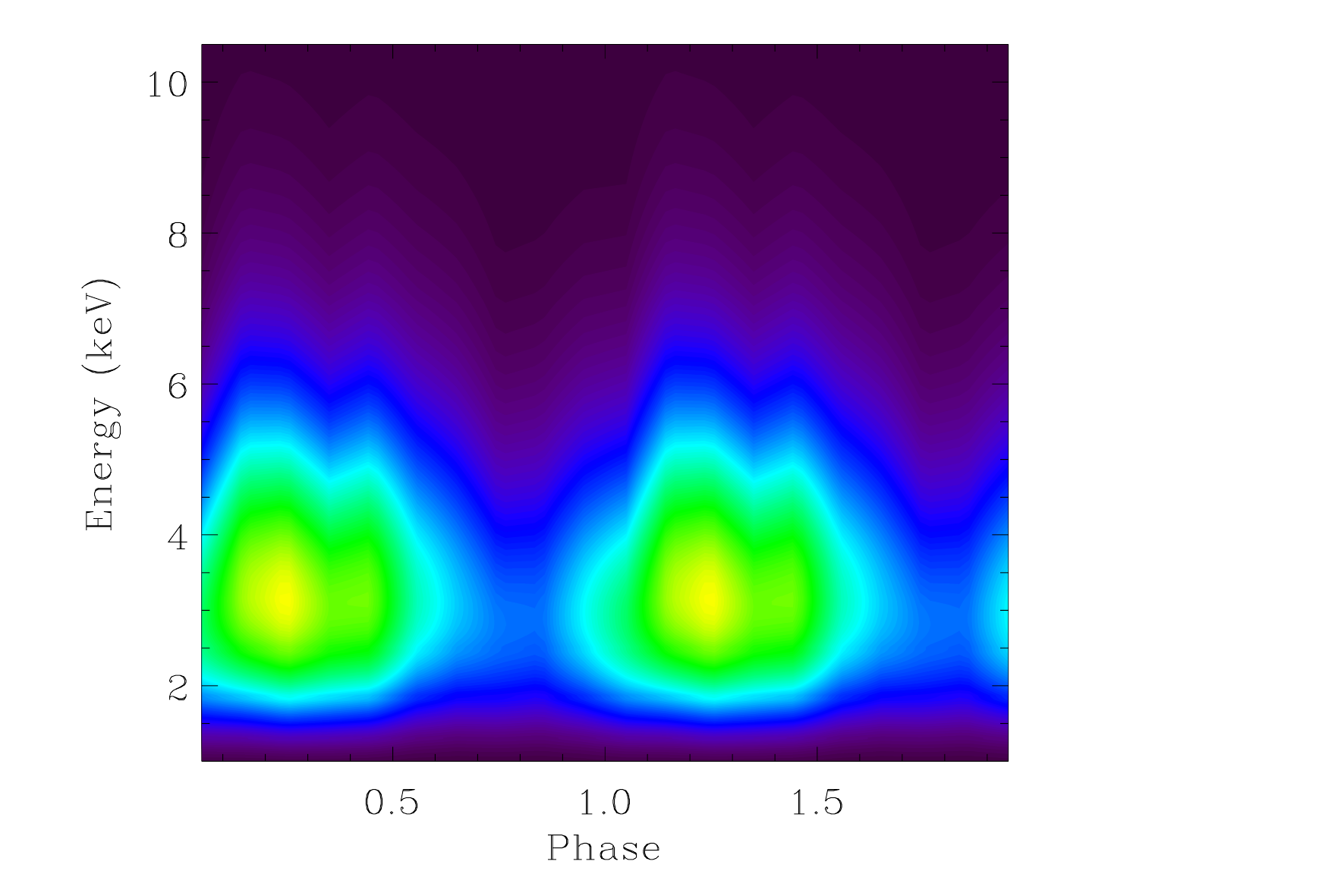}}
\vbox{
\includegraphics[height=2.5cm]{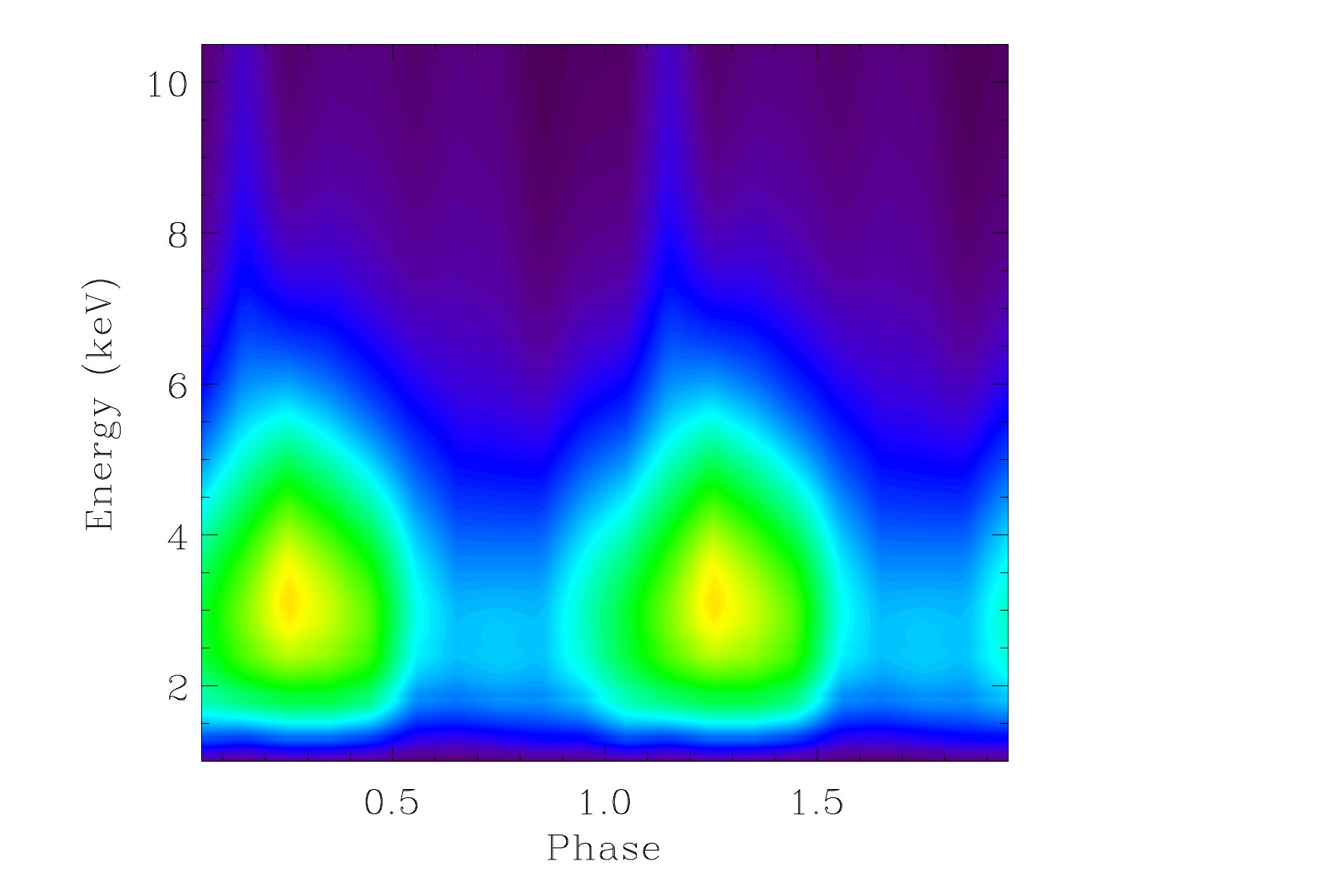}
\includegraphics[height=2.5cm]{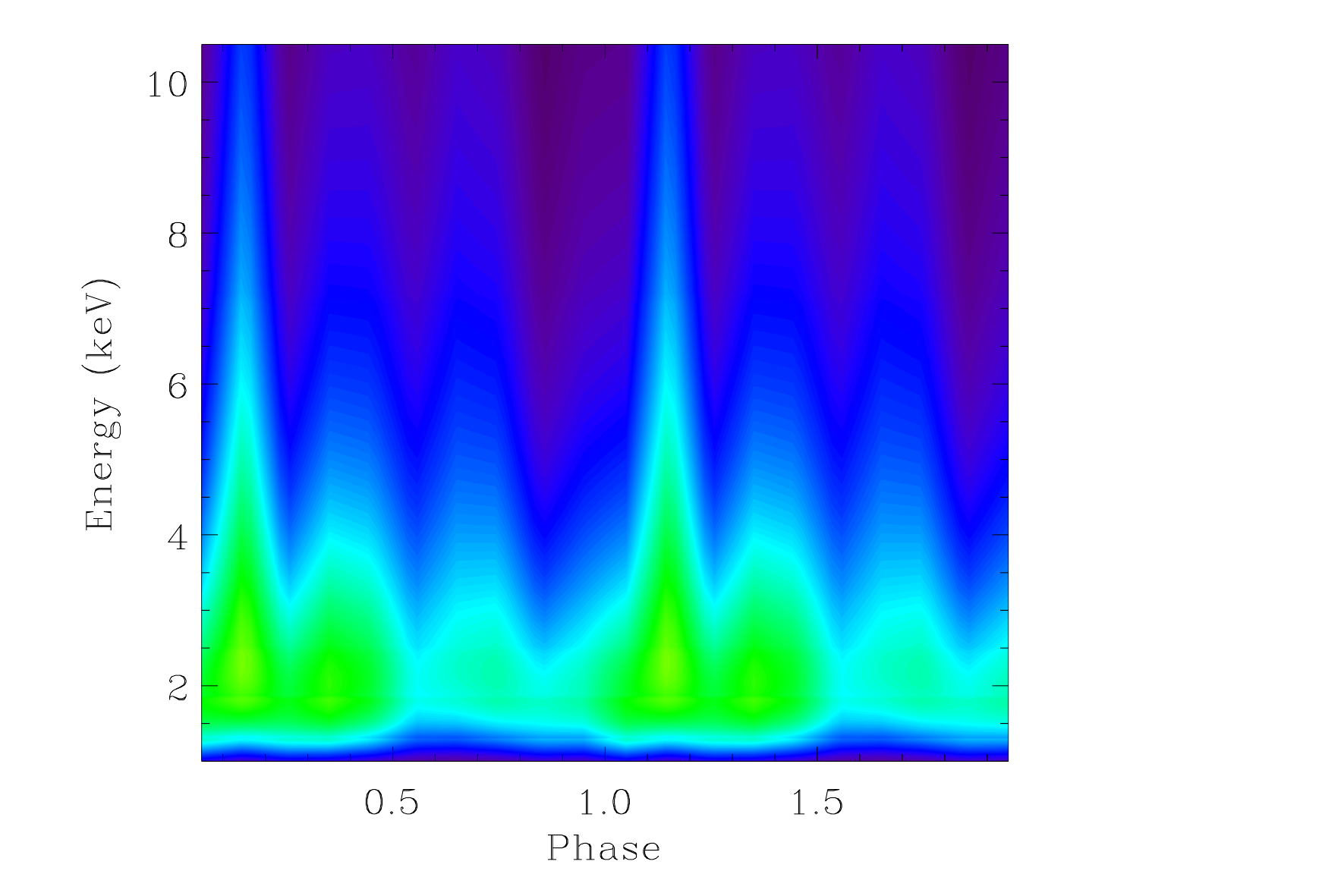}
\includegraphics[height=2.5cm]{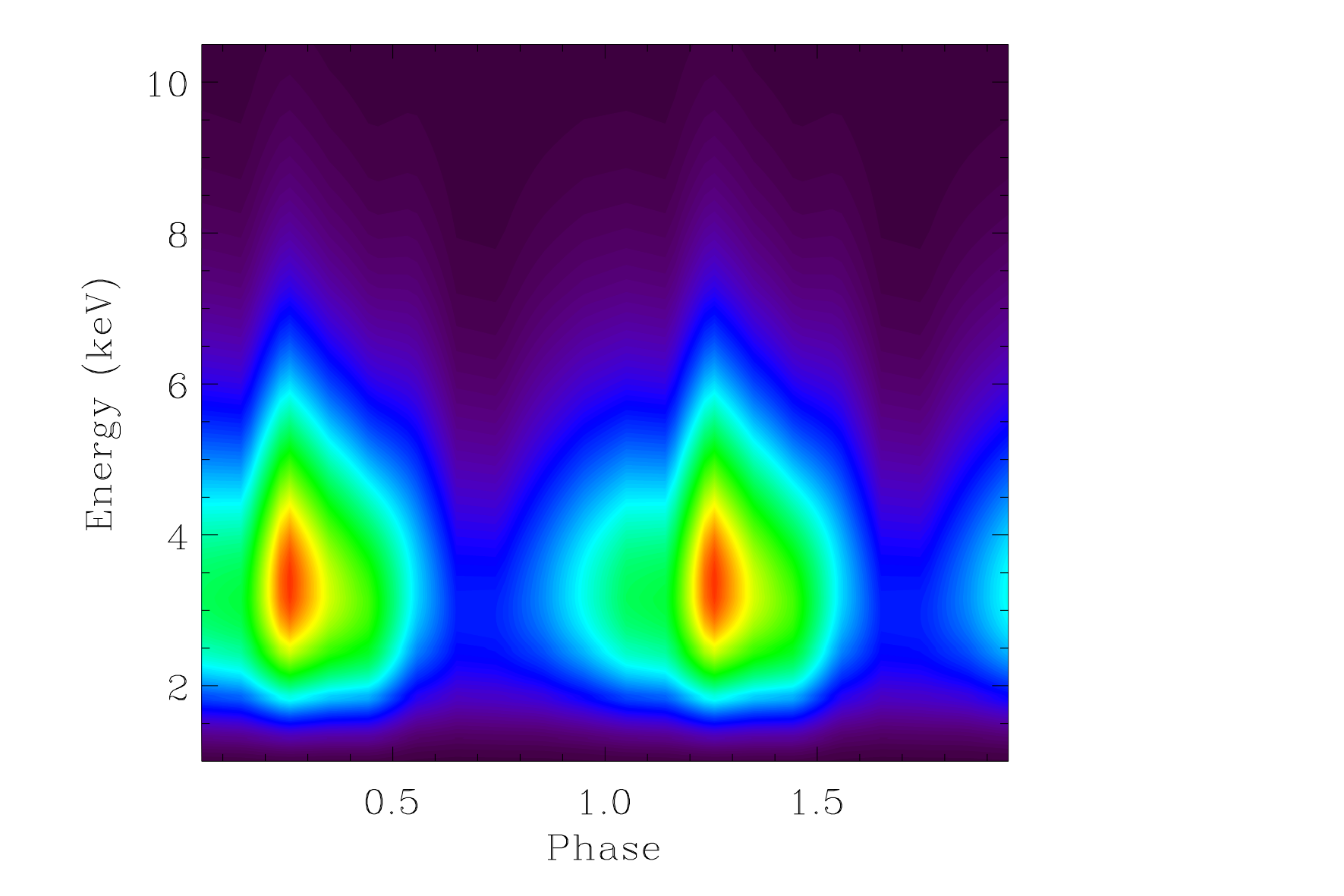}}
\vbox{
\includegraphics[height=2.5cm]{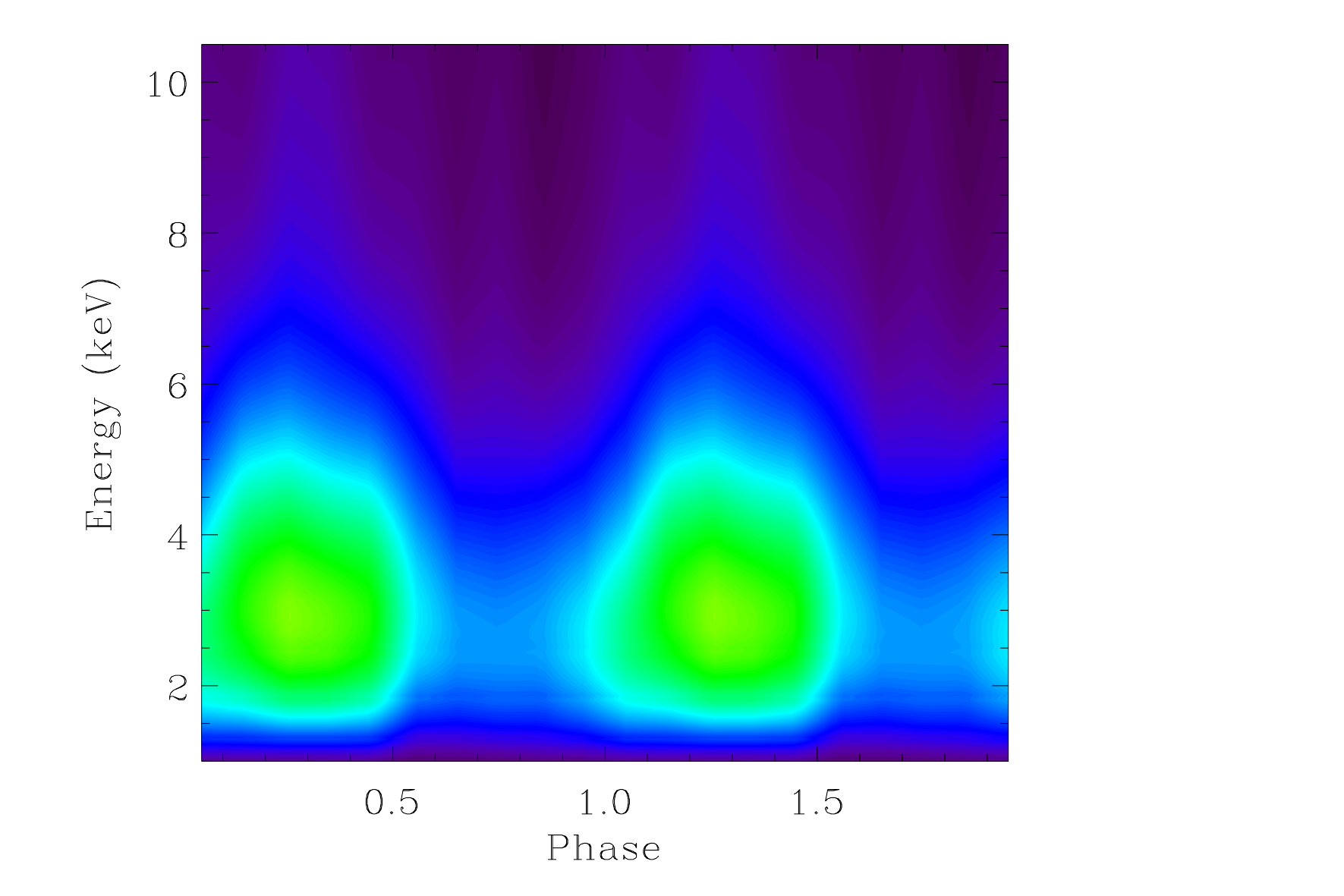}
\includegraphics[height=2.5cm]{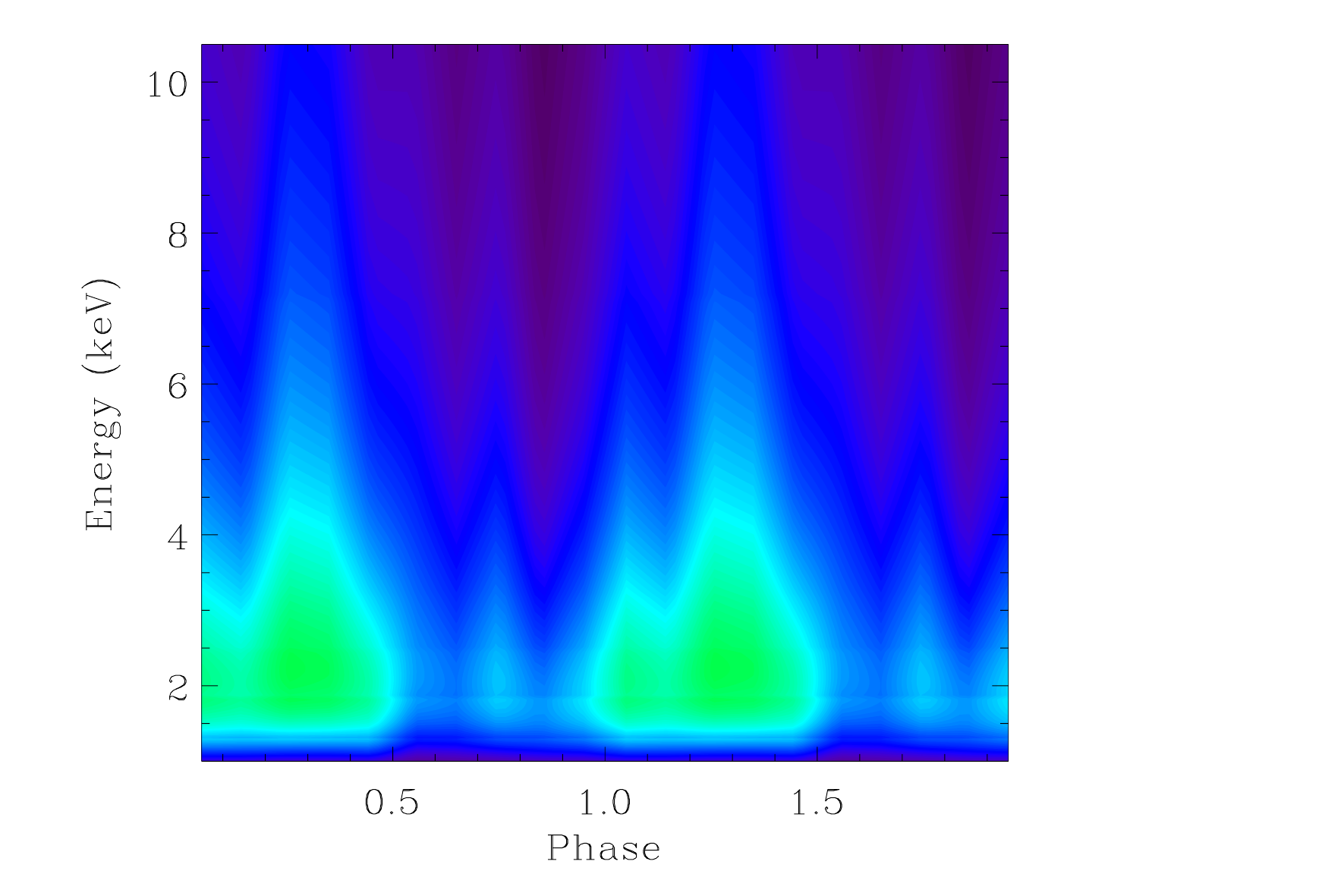}
\includegraphics[height=2.5cm]{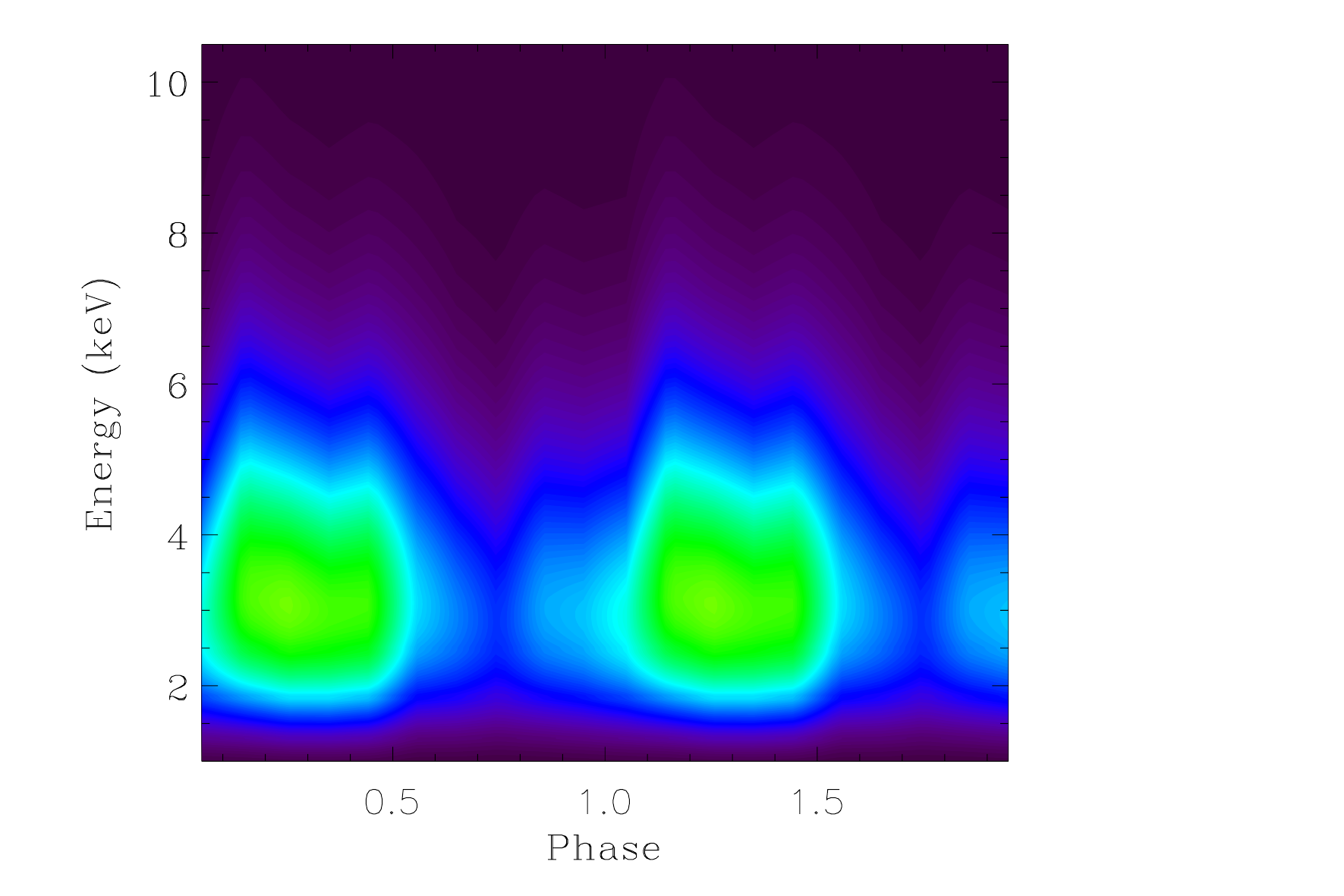}}
\vbox{
\includegraphics[height=2.5cm]{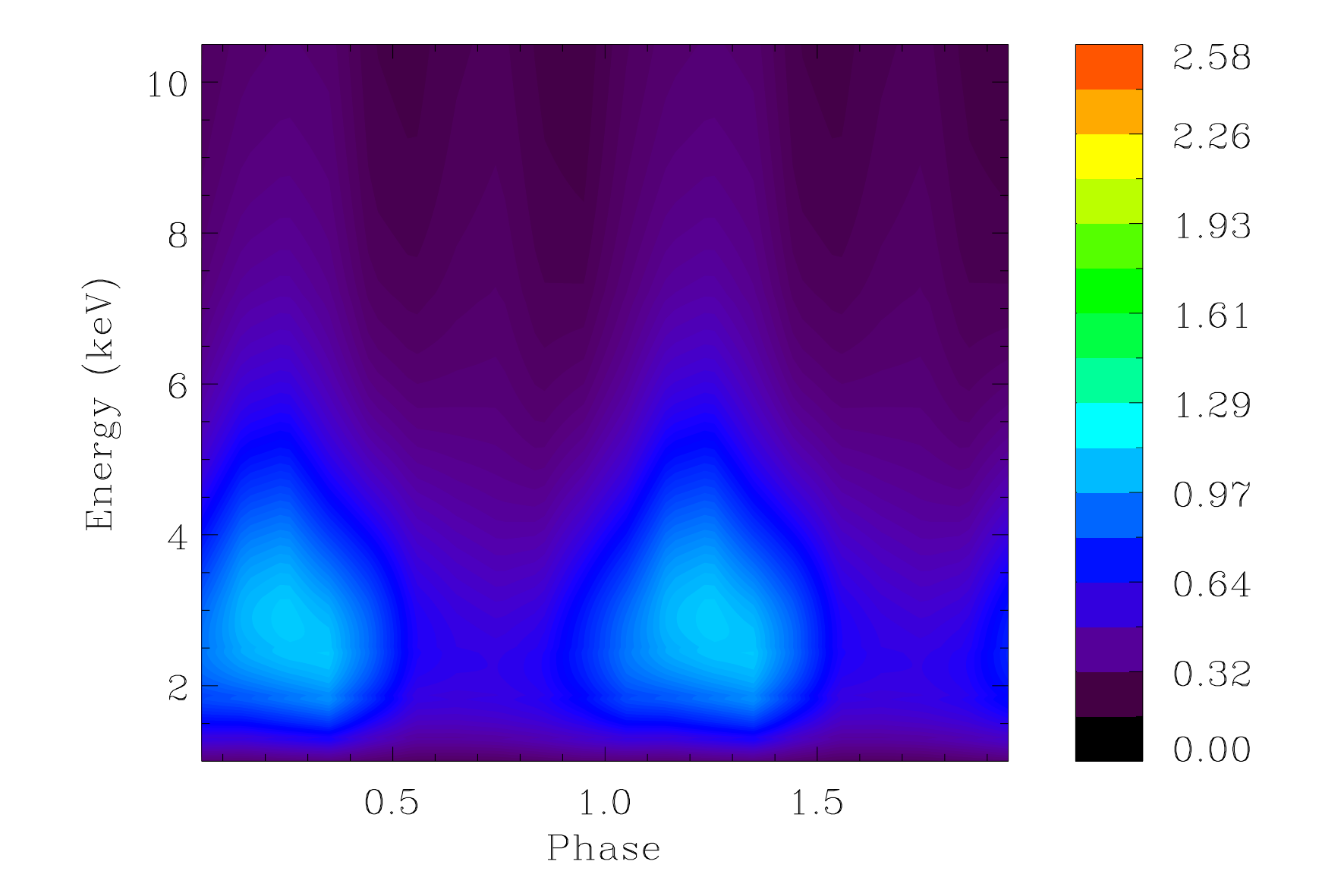}
\includegraphics[height=2.5cm]{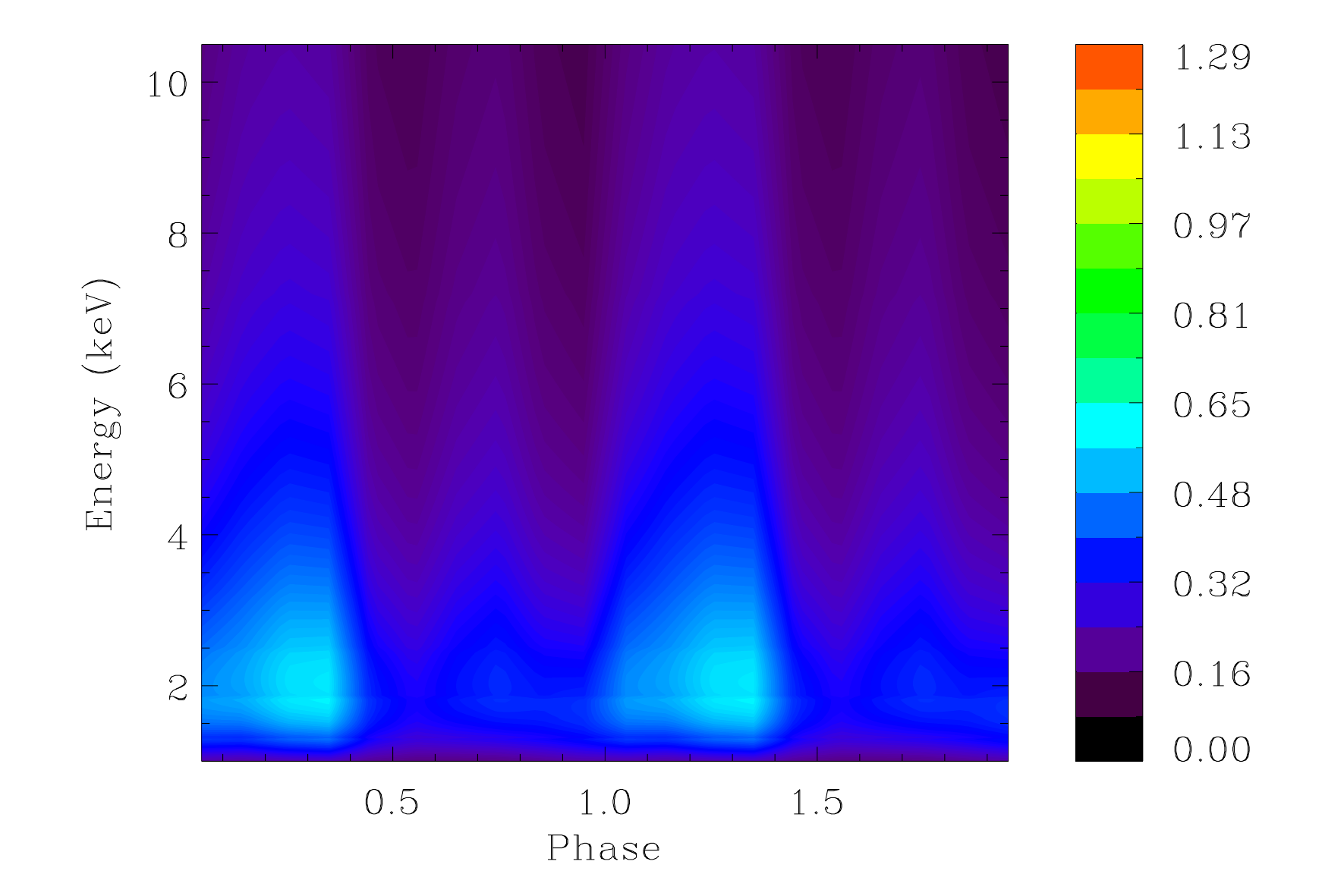}
\includegraphics[height=2.5cm]{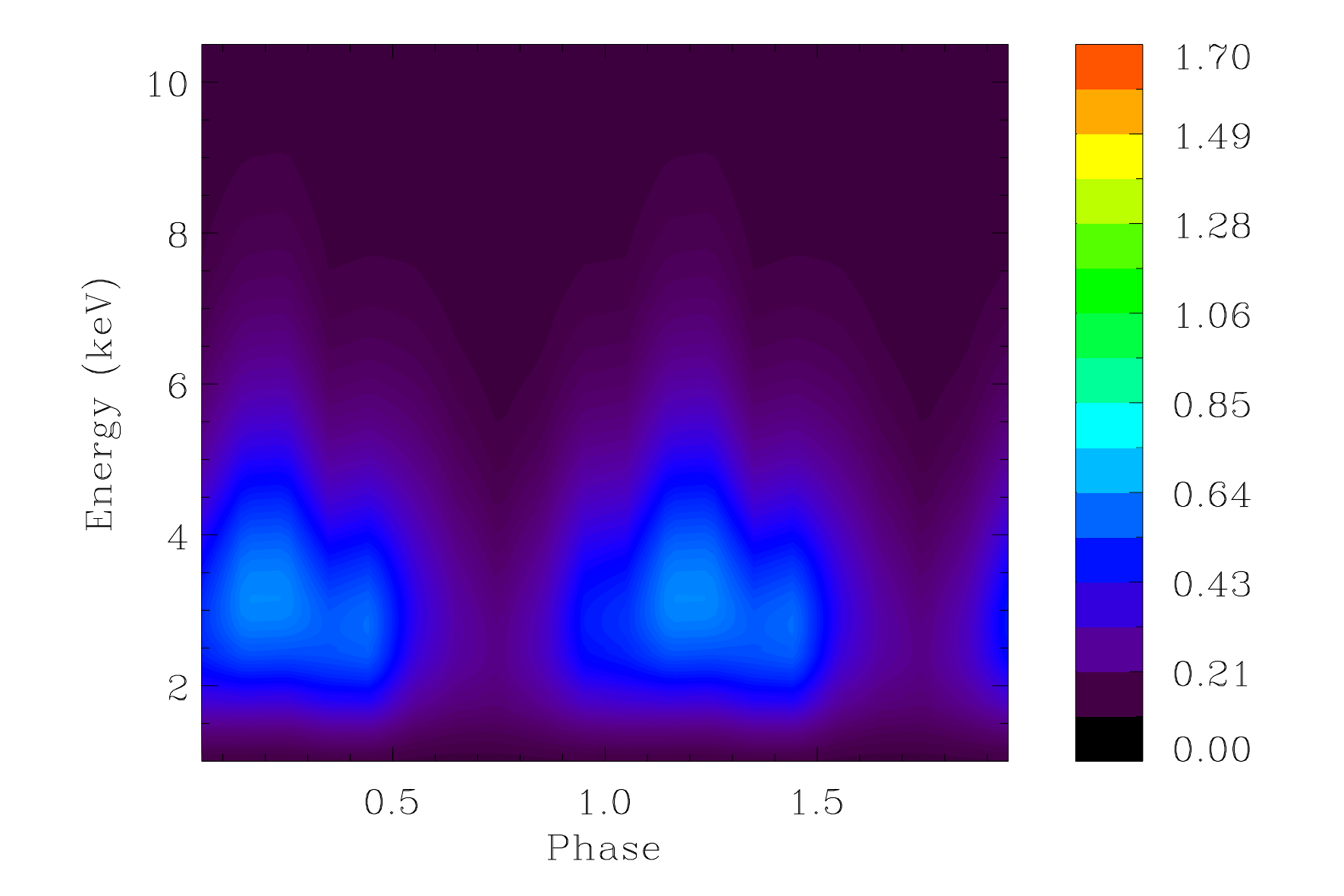}}

\caption{Dynamic Spectral Profiles (DSPs). Each row corresponds to
  one \xmm\, observation (epoch increases from top to bottom: 2008
  August 23, 29, 31, September 02 and 30) of \sgre\, \citep{rea09}. The three columns represent 
  in the phase/energy plane the contour plots for the total
 (left), power-law (middle) and
  blackbody (right) $\nu$F$_\nu$ flux. The color scale is in
  units of 0.01\,keV(keV\,cm$^{-2}$\,s$^{-1}$\,keV$^{-1}$).}
\label{dsp} 
\end{figure*}


\subsection{\sgre}

A new magnetar candidate, \sgre\, was discovered on 2008 August
22 by the \swift\, Burst Alert Telescope (BAT), thanks to the
detection of SGR-like bursts. Tens of  bursts were observed \citep{mereghetti09,enoto09,rea09,gogus10} with fluxes
exceeding the underlying continuum by a factor $>$$10^5$. The bursts reached a maximum luminosity of
$\sim$$10^{41}$\ergs and had a durations of $<$1\,s, typical of 
those usually emitted by SGRs. Thanks to the rapid response 
of many X-ray satellites (\swift, \su, \xmm, \CXO, \INT, Fermi-GBM, AGILE) 
the source was repeatedly observed during and after its ``burst forest'' emission,
leading to the best  monitoring of an SGR outburst ever performed.  

Archival ROSAT data dated September 1992 showed a faint unpulsed X-ray
source consistent with the position of this new SGR, most probably its
quiescent X-ray counterpart \citep{rea09}, at a flux level of $\sim
10^{-12}$~erg\,s$^{-1}$cm$^{-2}$, and with a thermal spectrum well modeled
by a blackbody with a temperature of kT$\sim$0.3\,keV (although a
power-law fit is also consistent with the data).

Since the very first \xmm\ observation of \sgre, carried out one day
after the source bursting activation, it was already clear that the source
spectral and timing properties changed enormously with respect to the
ROSAT quiescent level, with a flux increase of almost 2 orders of
magnitude and a much harder spectrum. The source characteristics
evolved rapidly during the outburst decay showing a significant
spectral softening in the first month of monitoring correlated with
the flux decay \citep{rea09}. The source rotate at $\sim$5.7\,s , and a period
derivative of $\dot{P}=6.8\times10^{-12}$ \ss . Its pulse profile is highly variable (in time and
as a function of energy), and it has a quite stable pulsed fraction of $\sim$40\%,

Given the good monitoring of this outbursts, a phase resolved decay has been observed for the first time. We show this in Figure \ref{dsp} . This clearly shows how different is the outburst evolution in phase, as well as the different decays of the thermal (much slower) and non-thermal (much faster) components.

\subsection{\sgrf}

SGR\,0418+5729 was discovered due to its bursting activity in 2009 June 5 \citep{vanderhorst10} and then its post-outburst behaviour was intensively monitored for about 160 days \citep{esposito10}. Only three bursts were detected from this SGR (all on 2009 June 5 \citep{vanderhorst10}) and archival searches could not reveal any previous period of intense activity \citep{vanderhorst10}; no new bursts have been detected until now. SGR\,0418+5729 has a spin period of 9.1 s and its period derivative has so far eluded all measurements during its outburst decay. 

After the outburst onset, the source spectrum gradually softened and the persistent X-ray emission faded by a factor of $\sim$10 in about 160 days following a broken-power-law decay, with a steepening about 19 days after the activation, when the index changed from $-0.3$ to $-1.2$ \citep{esposito10}. At the same time, the complex and energy-variable pulse profile of the source showed a substantial evolution, pointing to a nearly orthogonal rotator seen at a large inclination angle and to the presence of two emitting caps, one of which became hotter during the outburst \citep{esposito10}.

\begin{figure}[t]
\includegraphics[width=12cm]{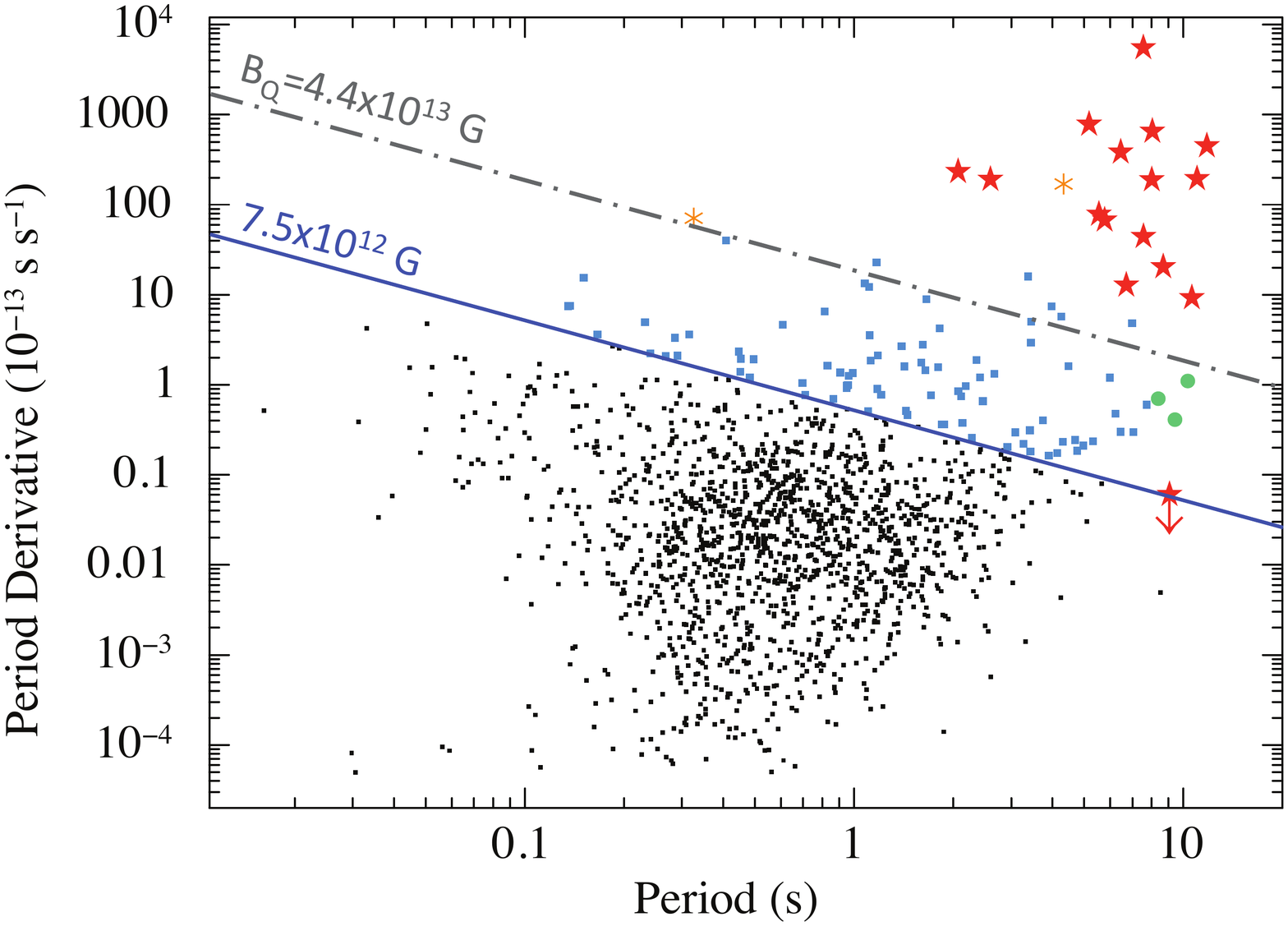}
\caption{Diagram of the $P$--$\dot{P}$ of all isolated pulsars known to date. Black squares represent normal radio pulsars, light-blue squares normal radio pulsars with a magnetic field larger than 7.5$\times10^{12}$\,Gauss (our limit for SGR 0418+5729), red stars are the magnetars, orange asterisks are the magnetar-like pulsars \hbpsr\, and \psr, and the green circles are the X-ray Dim Isolated Neutron Stars (XDINSs). The blue solid line marks the 90\% upper limit for the dipolar magnetic field of SGR 0418+5729. The value of the electron quantum magnetic field is also reported (dash-dotted grey line). See \citep{rea10} for more info. }
\label{ppdotsgr}     
\end{figure}

The most intriguing characteristic of this object comes from the current non-detection of its period derivative after $\sim$500\,days from the outburst onset. In particular, the 90\% limit on it is $< 6\times10^{-15}$ \ss , leading to an upper limit on its magnetic field of $B < 7.5\times10^{12}$\,Gauss: an incredibly low magnetic field for an SGR \citep{rea10}. This source shows that the magnetar population may thus include objects with a wider range of B-field strengths, ages and evolutionary stages than observed so far (see also Figure~\ref{ppdotsgr} and \citep{rea10,pp10}) .

\subsection{\sgrg}
This recent addition to the magnetar family was discovered on 2010 March 19, when the Swift triggered on a short hard X-ray burst and localized it in a region close to the Galactic plane \citep{gogus10}. Swift immediately slewed to the the BAT field and unveiled the existence of a previously unknown bright X-ray source. Given the proximity to the Galactic plane and the burst properties, the X-ray source was immediately suggested to be an SGR. Its SGR/magnetar nature has been confirmed shortly after by the discovery with Swift and RXTE of pulsations at 7.57 s and the measure of a period derivative of a few $10^{-12}$ s s$^{-1}$ \citep{gogus10,eit10}. 

Follow-up observations could not reveal SGR\,1833--0832 in the  optical, infrared, and radio wavebands \citep{gogus10,eit10}. The source was monitored in the X-rays with various instruments for about five months \citep{gogus10,eit10}, during which it displayed a low bursting activity, while the persistent flux decreased from $\sim$$4\times10^{-12}$ erg cm$^{-2}$ s$^{-1}$ by a factor of about four with an approximately exponential decay with a time scale ($e$-folding time) of $\sim$110 days \citep{eit10}. Somewhat atypically, the spectrum of this source is well described by an absorbed single hot blackbody ($kT\sim1.2$ keV). However, the high absorbing column towards the source makes it difficult to properly model its emission and an additional spectral component cannot be excluded \citep{eit10}.


\subsection{\axj}
This source was discovered in a 1993 observation of the (not associated) supernova remnant Kes~75 during a search for pulsating sources \citep{torii98,gotthelf98}. The observed flux was $\sim$$4\times10^{-12}$~erg~cm$^{-2}$~s$^{-1}$. The long period (7\,s) together with the spectral properties and the lack of a companion suggested that \axj\,  is an AXP. The source was observed again with ASCA in 1997 and 1999 and revealed only in the 1999 observation, at a ten times lower flux (which is consistent with the marginal 1997 non detection) \citep{vasisht00}. Subsequent BeppoSAX, Chandra, and XMM-Newton observations revealed an X-ray source at a flux level similar to that observed from AX\,J1845.0--0258 in 1999 \citep{israel04,tam06}. The low flux precluded new measurements of pulsations and therefore there is no information about the rate of change of the spin period (for this reason the source is often indicated as a candidate AXP).

 If \axj\,  is indeed an AXP, it is plausible that like other magnetars it has a transient behavior and that the 1993 ASCA observation was carried out during (or shortly after) an outburst. Moreover, Tam et al. \citep{tam06} noticed that, given the large uncertainty on the ASCA position, it is not ruled out that the X-ray source observed by the other satellites is a field-source unrelated to the 7-s pulsar \axj . If so, the upper limit on the flux that they obtained from a deep Chandra observation of the field would point to a flux decrease of a factor of 200 or more.

\section{Conclusions}

In this review we reported on all magnetar outbursts recorded to date (see also Figure \ref{outbursts}). Although far from having an exhaustive understanding on the mechanism and observational behavior of magnetar outbursts (see \citep{pp10} for a recent work in that direction), some common lines can be drawn. Before listing the few common properties as they appear up to know, we caveat that many bias might be in place. The most evident is that in several cases we have no information on the exact time of the outburst onset, hence the observations we are comparing for different sources might be relative to different times during the outburst decay. 

\begin{itemize}

\item A clear softening of the X-ray spectrum during the outburst decay is observed in all cases. There is a large amount of higher energy photons emitted at the beginning of the outburst that fade during the flux decay toward quiescence. In some cases this was even connected to the detection of a transient hard X-ray emission days after the on-set of the outburst, which faded away very rapidly ($\sim$ a week; see the case of \sgre ).

\item There is a connection between the enhancement of the persistent emission and the source bursting activity, although the sparse observations does not allow to figure out whether there is a connection between the fluence of those outbursts and the amount or energetic of the short bursts.

\item A single source can emit several outbursts (see e.g. the case of \aa ), of different intensities and timescales. 

\item There is no clear understanding on the radio, infrared and optical emission during outbursts. In the radio band, in some cases radio pulsed emission were detected in connection to the X-ray enhancements (it is not clear yet who triggers who, though), in others not. On the other hand, infrared and optical emission were sometimes found to get enhanced by the X-ray outburst activity, while in other cases they were not.

\item Large differences in the timescales and decay-laws are observed among different outbursts, even in the same source.

\end{itemize}

\begin{acknowledgement}

NR acknowledges support from a Ramon y Cajal Fellowship and from grants AYA2009-07391 and SGR2009-811. PE acknowledges financial support from the Autonomous Region of Sardinia through a research grant under the program PO Sardegna FSE 2007--2013, L.R. 7/2007 ``Promoting scientific research and innovation technology in Sardinia''. 

\end{acknowledgement}

\bibliographystyle{spmpsci}
\bibliography{rea_heeps_final.bbl}

\end{document}